\documentclass[twocolumn]{aastex62}
\usepackage[utf8]{inputenc}

\begin{document}

\title{A Revised Exoplanet Yield from the Transiting Exoplanet Survey Satellite (TESS)}

\author{Thomas Barclay}
\affiliation{NASA Goddard Space Flight Center, 8800 Greenbelt Road, Greenbelt, MD 20771, USA}
\affiliation{University of Maryland, Baltimore County, 1000 Hilltop Cir, Baltimore, MD 21250}

\author{Joshua Pepper}
\affiliation{Department of Physics, Lehigh University, 16 Memorial Drive East, Bethlehem, PA, 18015, USA}

% \author{Joshua E. Schlieder}
% \affiliation{NASA Goddard Space Flight Center, 8800 Greenbelt Road, Greenbelt, MD 20771, USA}

\author{Elisa V. Quintana}
\affiliation{NASA Goddard Space Flight Center, 8800 Greenbelt Road, Greenbelt, MD 20771, USA}

% \date{\today}

\begin{abstract}
The Transiting Exoplanet Survey Satellite (TESS) has a goal of detecting small planets orbiting stars bright enough for mass determination via ground-based radial velocity observations. 
Here we present estimates of how many exoplanets the TESS mission will detect, physical properties of the detected planets, and the properties of the stars that those planets orbit. This work uses stars drawn from the TESS Input Catalog Candidate Target List and revises yields from prior studies that were based on Galactic models. We modeled the TESS observing strategy to select approximately 200,000 stars at 2-minute cadence, while the remaining stars are observed at 30-min cadence in full-frame image data. We placed zero or more planets in orbit around each star, with physical properties following measured exoplanet occurrence rates, and used the TESS noise model to predict the derived properties of the detected exoplanets. In the TESS 2-minute cadence mode we estimate that TESS will find $1250\pm70$ exoplanets (90\% confidence), including 250 smaller than 2 Earth-radii. Furthermore, we predict an additional 3100 planets will be found in full-frame image data orbiting bright dwarf stars and more than 10,000 around fainter stars. We predict that TESS will find 500 planets orbiting M-dwarfs, but the majority of planets will orbit stars larger than the Sun. Our simulated sample of planets contains hundreds of small planets amenable to radial velocity follow-up, potentially more than tripling the number of planets smaller than 4 Earth-radii with mass measurements. This sample of simulated planets is available for use in planning follow-up observations and analyses.

\end{abstract}

\section{Introduction}
While we have known that planets orbit stars other than the Sun since the late 20th Century \citep{Walker1988,Latham1989,Wolszczan1992,Mayor1995}, it is only with the launch of the Kepler spacecraft in 2009 \citep{Koch2010,Borucki2010} that we have been able to estimate the occurrence rates of terrestrial worlds. While there is not a firm consensus on the details of how common planets are as a function of size and orbital period \citep{Howard2010,Gould2010,Catanzarite2011,Youdin2011,Howard2012,Traub2012,Bonfils2013,Swift2013,Fressin2013,Petigura2013a,Petigura2013b,Montet2014,Kane2014,Foreman2014,Burke2015,Clanton2016,Hsu2018} it is clear that exoplanets overall are fairly commonplace, particularly orbiting the coolest of stars \citep{Dressing2013,Dressing2015,Morton2014,Mulders2015}.

Although we have a fairly large sample of planets with orbital periods of less than a few hundred days, there is still a pressing need to detect planets that are readily characterizable. The primary goal of the Transiting Exoplanet Survey Satellite (TESS), a mission led by the Massachusettes Institute of Technology, is to find small planets that are most amenable for mass measurements through precise radial velocity observations \citep{Ricker2015,Ricker2016,Collins2018}. A secondary, although unofficial, mission goal is to find targets that can be characterized through transmission spectroscopy from the James Webb Space Telescope and other future observatories.

\begin{figure}
\includegraphics[width=0.45\textwidth]{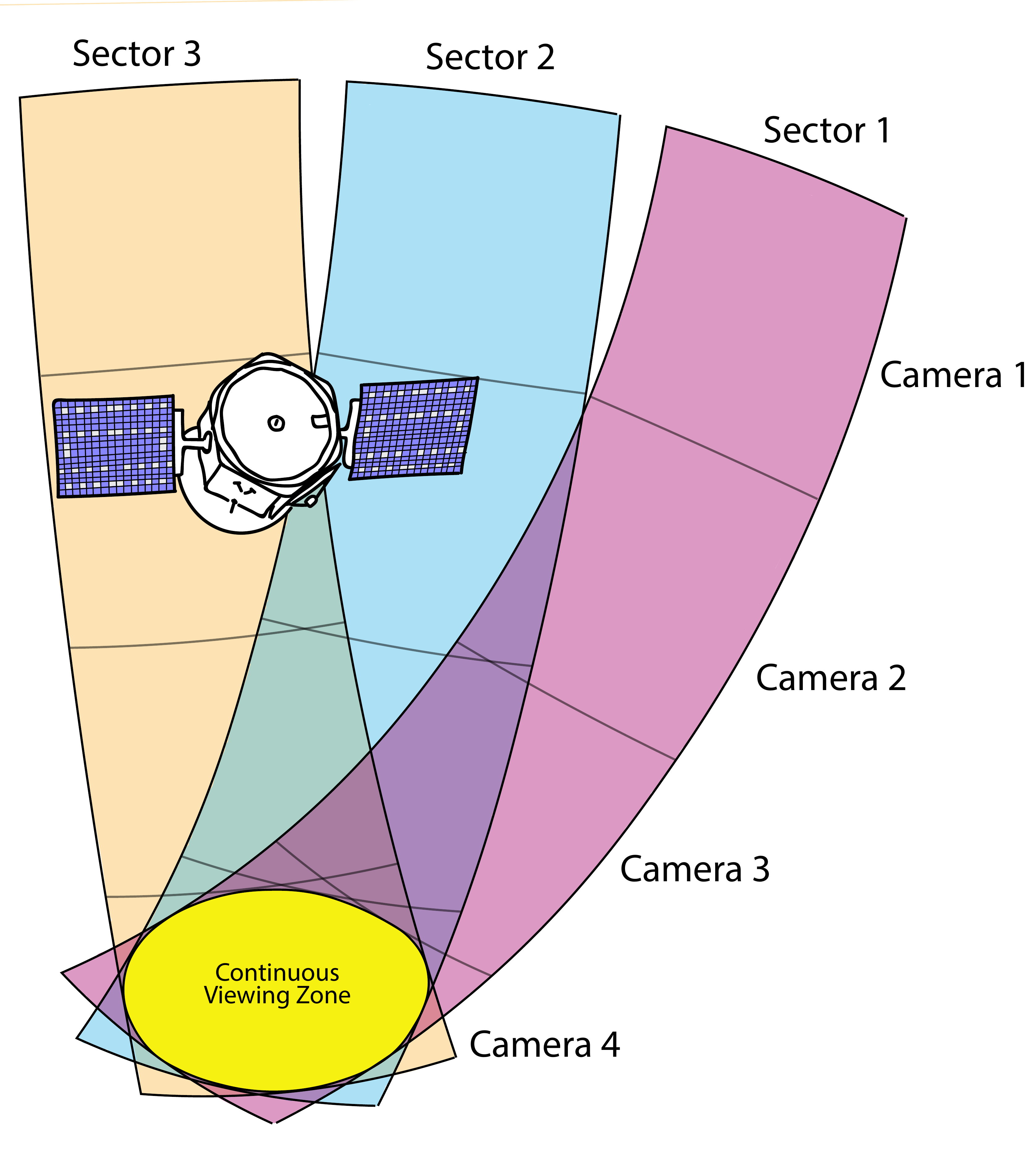}
\caption{An illustration showing the first three sectors of the TESS observing plan.}
\label{fig:sectors}
\end{figure}

TESS launched on April 18, 2018, and resides in an elliptical 13.7 day high-Earth orbit during a 2-year primary mission. TESS has four cameras, each with a $24^{\circ}\times24^{\circ}$ field of view. The cameras are aligned to provide continuous coverage of $96^{\circ}\times 24^{\circ}$, which is maintained for 27.4 days per pointing (known as a sector). The long axis of the observing region is aligned with a fixed ecliptic longitude, with the boresight of the fourth camera centered on the ecliptic pole, as shown in Figure~\ref{fig:sectors}. Every two orbits, TESS rotates $\sim$28$^{\circ}$ about the ecliptic pole. In year 1 of the mission, the spacecraft will survey 13 sectors in the southern ecliptic hemisphere, before spending year 2 in the northern ecliptic hemisphere. About 60\% of the sky will be covered by a single sector of TESS observations, and a further 15\% will be observed over two sectors, located in the overlap areas between two adjacent sectors. Most stars within 12 degrees of the ecliptic poles will be within the TESS continuous viewing zone (CVZ) and observable for more than 300 days (this accounts for approximately 1\% of the sky per pole). Over the course of the prime mission, TESS will observe approximately 85\% of the sky.

The TESS mission is focused on detecting small transiting planets that orbit bright stars. Although the dwell time over most of the sky is too short to permit the detection of planets in temperate orbits, that goal can be advanced by discovering planets orbiting cooler stars, especially in the TESS CVZ around the ecliptic poles. 

Two observing modes will be initially implemented: the $96^{\circ} \times24^\circ$ full-frame image (FFI) will be recorded every 30-minutes, while approximately 200,000 stars will be preselected to have data recorded at 2-minute cadence. In either case, the system is integrating and reading out every 2 seconds; they differ in the number of coadds.

It is essential that a \emph{reasonable} prediction for the scientific yield of TESS is available because (a) planning follow-up resources requires knowing the properties of the planets we might find \citep{Louie2018,Crouzet2017,Collins2018,Kempton2018}, (b) we can perform trade studies on target prioritization schemes for the 2-minute cadence targets \citep[][Pepper et al. in preparation]{Bouma2017,Stassun2017}, and when designing data analysis algorithms \citep{Kipping2017,Lund2017,pyke3,Oelkers2018}, and (c) we can manage the expectations of the scientific community and the public.

A TESS yield simulation created by \citet{Sullivan2015} has been the standard used by both the mission team and the community. Since then, two papers have built on the work of \citeauthor{Sullivan2015} to refine the total mission yield and explore extended mission scenarios \citep{Bouma2017}, and to improve estimates of the planet yield from M-dwarfs \citep{Ballard2018}. However, \citet{Sullivan2015} simulations were based on a simulated stellar population rather than real stars, and used an earlier hardware configuration that provided for greater storage and downlink limits than the flight hardware being used. Therefore, now is the time to revise the TESS yield estimate using new information. Here we report on a new estimate of the exoplanet yield using the TESS Input Catalog (TIC) Candidate Target List (CTL), the same list that is used by the mission to select stars and perform photometry.

% ---------------

\section{Simulating stars, planets, and detections}

The process we used to derive a population of planets detectable by TESS uses a Monte Carlo method to (1) simulate the population of stars that TESS will observe, (2) place planets in orbit around these stars, and (3) predict how many of these planets TESS will detect.

\subsection{Star selection}
\label{sec:starselction}
The first step was made relatively straightforward by the availability of the CTL - a prioritized list of target stars that the TESS Target Selection Working Group have determined represent the stars most suitable for detection of small planets by TESS. The properties of about 500 million stars were assembled in the TIC \citep{Stassun2017}, and the CTL includes several million of those stars that are most suitable for small transit detection. We used CTL version 6.1\footnote{The TIC and CTL are available from the MAST archive at \url{http://archive.stsci.edu/tess/}.}, which includes 3.8 million stars with properties such as stellar radii, masses, distances, and apparent brightness in various bandpasses. The CTL stars were then ranked using a simple metric based on stellar brightness and radius, along with the degree of blending and flux contamination (especially important given the large TESS pixels). The CTL does not include all stars. Save for stars on specially curated target lists \citep[e.g.][]{Muirhead2018}, stars with reduced proper motions that indicate they are red giants \citep{Collier2007}, stars with a temperature below 5500 K and a TESS magnitude fainter than 12, or stars with temperature above 5500 K and a TESS magnitude fainter than 13, are excluded from the CTL. Such broad cuts were required in order to assemble a small enough population of stars to practically manage.

We then determined which of these stars are likely to be observed by the mission. We used {\it tvguide} \citep{Mukai2017} on each star to determine whether and for how long it is observable with TESS. We arbitrarily selected a central ecliptic longitude for the first sector of 277$^\circ$ which equates to an antisolar date of June 28 (the precise timing of the first sector is dependent on commissioning duration). Until we have on-orbit measurements of focal plane geometry, {\it tvguide} assumes that the cameras are uniform square detectors projected on the sky, placed precisely 24$^\circ$ apart in ecliptic latitude and with identical ecliptic longitude. Gaps between CCDs are assumed to be 0.25$^\circ$. We ended up with a total of 3.18 million individual stars on silicon. 

We also needed to simulate which of these stars are likely to be observed at 2-minute cadence and ensure compliance with the TESS mission requirement that states that over the 2-year mission over 200,000 total stars should be targeted, and 10,000 stars should be observed for at least 120 days. It is somewhat less trivial than one would initially assume to simulate this requirement because we could not simply select the top 200,000 stars with the highest priority in the CTL because this would place far too many stars in the CVZ than can actually be observed there at 2-minute cadence. To ensure a realistic distribution of targets, we first divided each ecliptic hemisphere into 15 sections: a polar section with everything within 13 degrees of the pole, representing stars that primarily fall into Camera 4; an ecliptic section including everything within 6 degrees of the ecliptic to represent stars that are not observed in the prime mission; and then the remaining area was divided longitudinally into 13 northern and 13 southern adjacent sections, representing stars observable with Cameras 1--3 in Sectors 1--26. This yielded a total of 28 sections of the sky with observable stars. 

A star that fell in a camera overlap region is observed in multiple sectors but only represented one unique target. We found that we could make a reasonable approximation to satisfy the requirements of 200,000 unique targets if in each polar section we selected the 6,000 stars in that region with the highest priority in the CTL, and then for each longitudinal section (representing the footprint of Cameras 1--3 in each sector) we selected the 8,200 highest priority stars in each of the regions. After removing stars that fall into CCD and camera gaps, this yielded 214,000 unique stars. We assumed that any star in an overlap region is observed in every possible sector.

\begin{figure}
\includegraphics[width=0.45\textwidth]{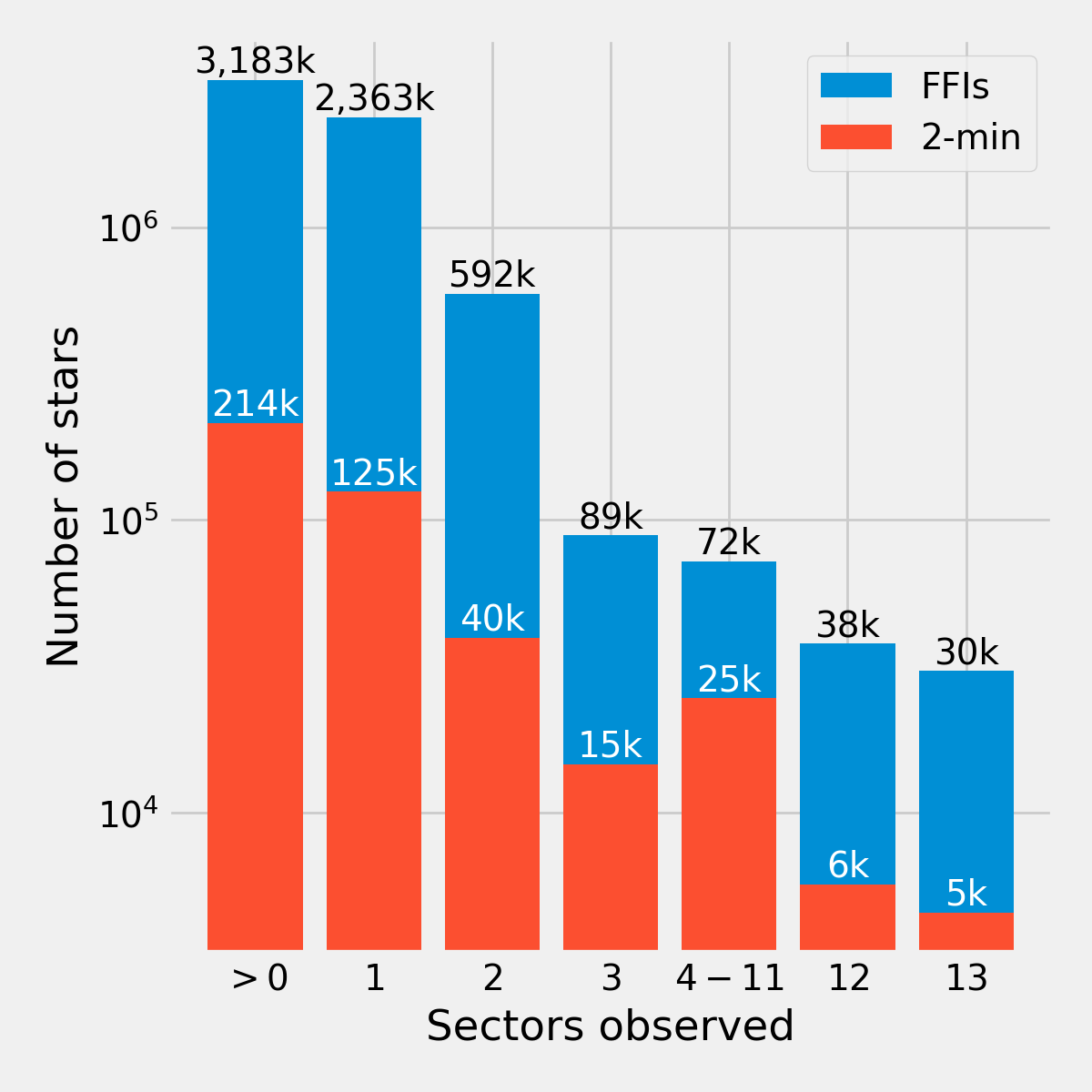}
\caption{The number of CTL targets observed for a given number of 27.4-day sectors. FFI targets are shown in blue, and 2-minute cadence targets in red. In total 3.2M CTL targets are observed, of which 214,000 are observed at 2-minute cadence. Roughly three-quarters of targets are only observed for a single sector, with just 2.1\% having 12 or 13 sectors of coverage. The 2-minute cadence targets are disproportionately observed for more sectors, with 4.2\% of the 2-minute cadence targets receiving 12 or 13 sectors of coverage.}
\label{fig:stars-sectors}
\end{figure}

While the CTL includes a great deal of curation, it is not infallible. A particular weakness inherent to stellar catalogs based on photometric colors is in distinguishing between dwarf stars and subgiants \citep{Huber2014,Mathur2017}. CTL versions up through 6.2 use parallax information when available to determine stellar radii (and therefore luminosity class), but the vast majority of stars depend on the use of reduced proper motion (RPM) cuts to distinguish dwarfs from giants. While GAIA DR2 will shortly provide reliable parallaxes for most CTL stars \citep{Huber2017,Davenport2017,Stassun2018}, the CTL will not be significantly modified until 2019. Furthermore, while the RPM method is highly reliable at distinguishing dwarfs and subgiants as a group from giant stars, it is generally not useful for distinguishing dwarfs from subgiants. Of the CTL stars that are classified as dwarfs based on the RPM cut, about 40\% are actually subgiants, although roughly 35\% of the CTL stars have parallax measurements confirming their spectral class. To account for this effect, we simulated a misclassified population of subgiants by increasing the stellar radius of 40\% of those AFGK stars which had been selected with the RPM cut by a factor 2, with the affected stars drawn at random. That included 25\% of all the AFGK stars in the CTL. This approach somewhat overestimates the radii of A-type subgiants but the effect on total planet yield is limited, because A-type stars have large radii, making detecting transiting planets challenging, and thus are already a relatively small fraction of the high-priority CTL stars.

\subsection{Simulating planets}
To each star in our list we assigned zero or more planets. The number of planets assigned to each star was drawn from a Poisson distribution. The mean (referred to here as $\lambda$) of the Poisson distribution we used differs between AFGK-dwarf stars and M-dwarfs because there is strong evidence that M-dwarfs host more planets on short orbital periods \citep{Mulders2015,Burke2015}. For AFGK stars we used the average number of planets per star with orbital periods of up to 85 days of $\lambda=0.689$ \citep{Fressin2013}, while for M-stars $\lambda=2.5$ planets are reported with orbital periods up to 200 days \citep{Dressing2015}.

Each planet was then assigned six physical properties drawn at random: an orbital period ($P$), a radius ($R_p$), an eccentricity, a periastron angle, an inclination to our line of sight ($i$), and a mid-time of first transit. The orbital period and radius were selected using the exoplanet occurrence rate estimate of \citet{Fressin2013} for AFGK stars, and \citet{Dressing2015} for M-stars. Both \citet{Fressin2013} and \citet{Dressing2015} reported occurrence rates in period/radius bins. We drew at random from each of these bins with the probability to draw from a given bin weighted by the occurrence rate in that bin divided by the total occurrence rate of planets. For example, \citet{Dressing2015} reported a 4.3\% occurrence rate for planets with radii 1.25--2.0 $R_\oplus$ and orbital period 10--17 days, so in our simulation we drew planets from that bin with a frequency of 4.3 divided by the total occurrence rate in all bins. We normalized by the total occurrence rate of planets since we already took account of systems with zero or multiple planets in the Poisson draw. Once we knew which bin to select a planet from, we drew from a uniform distribution over the bin area to select an orbital period-radius pair, except for the giant planet bin where we draw from a power-law distribution in planet radius with exponent -1.7, which mirrors \cite{Sullivan2015}. This non-uniform giant planet size distribution reduces the number of nonphysical inflated planets, as discussed by \citet{Mayorga2018}.
Occurrence rates from both \citet{Fressin2013} and \citet{Dressing2015} are based on Kepler data and are limited in orbital period to 0.5--85 and 0.5--200 days, respectively.

Following \citet{Kipping2014}, the orbital eccentricity was selected from a Beta distribution, with parameters $\alpha=1.03$ and $\beta=13.6$, which \citet{Vaneylen2015} found was appropriate for transiting planets. The periastron angle was drawn from a uniform distribution between $-\pi$ and $+\pi$. The cosine of inclination was chosen to be uniform between zero and one. Planets in multiple-planet systems were assumed to be coplanar - i.e. they have the same $\cos{i}$ - which is a reasonable appumption because multiple-exoplanet systems have been found to be highly coplanar \citep{Xie2016}. Finally, we chose a time of first transit to be uniform between zero and the orbital period -- note that this may be greater than the total observation duration, in which case no transit was recorded. We then computed the number of transits observed using the observation duration calculated previously (the number of sectors where a target is observed).

We intentionally kept planets that cross the orbit of other planets in the system because, while they are likely on unphysical orbits, to remove them would change the distribution of the number of planets per star, which is an observed property. We also assumed that none of these planets experience a significant amount of transit timing variations (Hadden et al. in preparation, address transit timing variations and period ratios in detail).

\subsection{Detection model}
Armed with a sample of planets and host stars, we then determined which planets are detectable. To do this we derived a transit depth modified by several factors: the flux contamination of nearby stars, the number of transits, and the transit duration. It should be noted that flux contamination is significantly more problematic for TESS than with Kepler because TESS has pixels that are 28 times larger than Kepler's.

The raw transit depth was computed assuming a uniform disk (i.e, transit depth $T_d = (R_p / R_\star)^2$, where $R_\star$ is the stellar radius). That is, we ignored the effects of limb-darkening and grazing transits. We calculated the reduction in transit depth due to dilution from nearby stars using the value of contamination for the CTL as $T_d / (1 + d)$, where $d$ is the dilution, the fraction of light coming from stars that are not the target divided by the total star light. We then multiplied the transit depth by the square root of the transit duration ($T_{\textrm{dur}}$) in hours, with transit duration following \citet{Winn2010} defined as,
\begin{equation}
    T_{\textrm{dur}} = \frac{P}{\pi} \arcsin\left[{ \frac{R_\star}{a} \frac{\sqrt{(1+R_p/R_\star) - b^2}}{\sqrt{1 - \cos^2{i}}}}\right],
\end{equation}
where $P$ is the orbital period, $i$ is the orbital inclination relative to our line of sight, $a/R_\star$ is the semimajor axis in units of stellar radius, $b$ is the impact parameter, and $R_p / R_\star$ is the planet to star radius ratio, to derive an effective transit depth.
The effective transit depth, $T_d^\prime$, is defined as
\begin{equation}
T_d^\prime = (R_p / R_\star)^2 \times \sqrt{T_{\textrm{dur}}} \times \sqrt{N} \times \frac{1}{1 + d},
\end{equation}
where $N$ is the number of transits observed.

We took the TESS photometric noise level from \cite{Stassun2017} who used the properties described by \citet{Ricker2016} and tested whether the effective transit depth was greater than the TESS photometric noise at the stellar brightness of the host stars multiplied by 7.3 (i.e. SNR$\ge$7.3). A 7.3-sigma detection is the nominal value used by \citet{Sullivan2015} and is calculated in a similar manner to the detection threshold used by Kepler \citep{Jenkins2010}. We also required that the impact parameter of the transit is less than 1.0 and that we observed at least 2 transits. Requiring an impact parameter of less than 1 removes a small number of grazing transits but these are difficult to distinguish from eclipsing binaries anyway \citep{Armstrong2017}. These detection thresholds are relatively aggressive, Section~\ref{sec:conservative} describes using a more conservative detection thresholds of at least 3 transits and SNR of 10.

% ---------------

\section{Results}
\label{sec:results}
\begin{figure}
\includegraphics[width=0.45\textwidth]{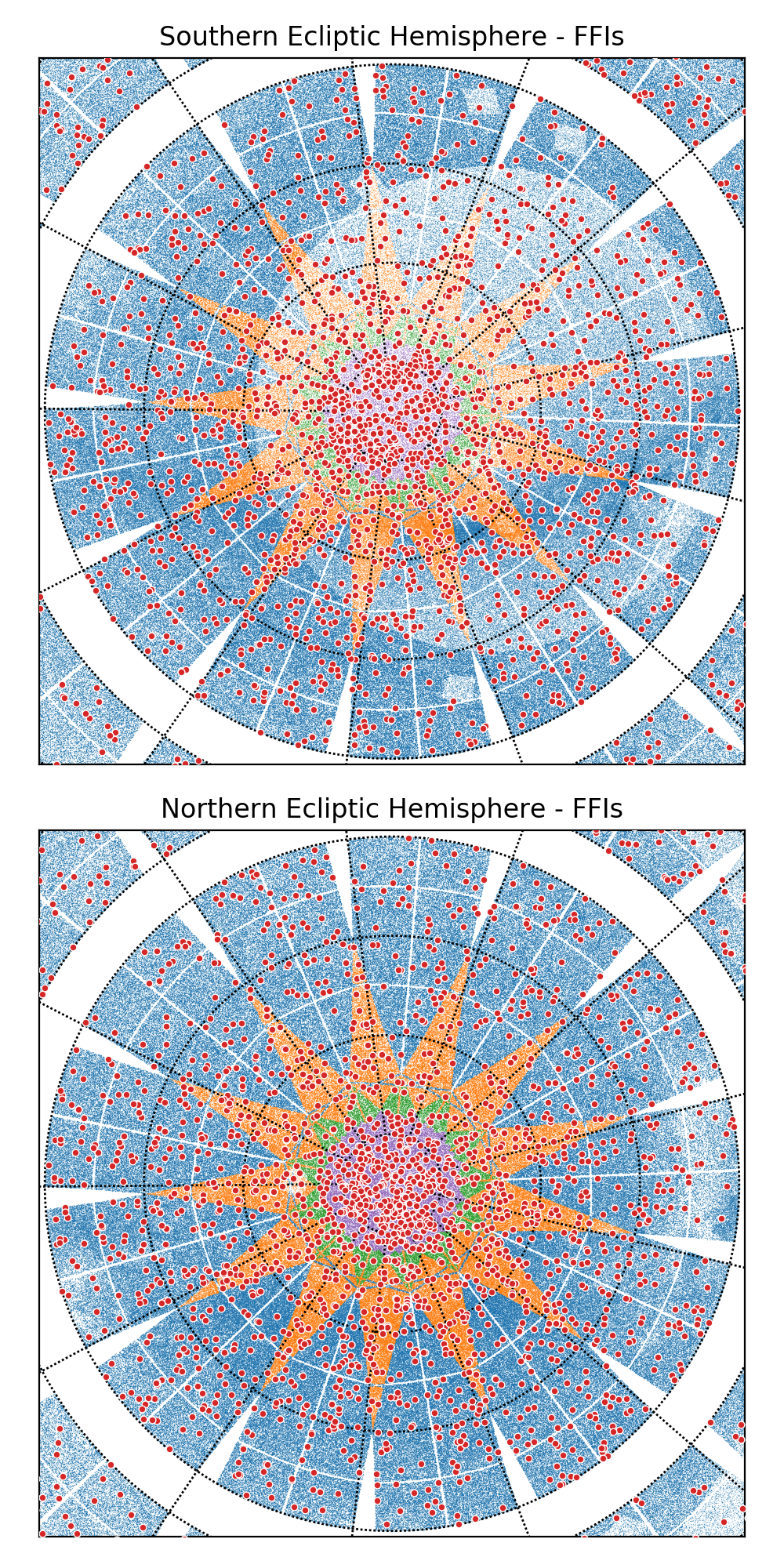}
\caption{The spatial distribution of target stars and detected planets from FFI data. The upper panel shows the southern ecliptic hemisphere and the lower panel shows the northern ecliptic hemisphere. Stars observed for 1 sector are shown in blue, two sectors in orange, 3+ sectors in green, and stars in the CVZ are shown in purple. Detected planets are shown as red dots. A total of 4373 planets are shown, of which 54\% were only observed for a single sector, and 11\% were observed for 12 or 13 sectors. The lower density of stars, offset from the south ecliptic pole, is centered on the south celestial pole, and is due to relatively incomplete proper-motion catalogs in the celestial south.}
\label{fig:skyplot-ffis}
\end{figure}

We performed 300 simulations using our nominal planet sample and detection criteria, this enabled us to look at the average and range from our simulations. We predict that TESS will find 4373 planets (median) orbiting stars on the CTL, with the 90\% confidence interval ranging from 4260--4510 planets. Henceforth, we designate a simulation that produced the median number of planets as our fiducial simulation and the properties we show come from that simulation. All the stars in the CTL are included in Figure~\ref{fig:skyplot-ffis} and the detected planets are shown as red dots. 

\begin{figure}
\includegraphics[width=0.45\textwidth]{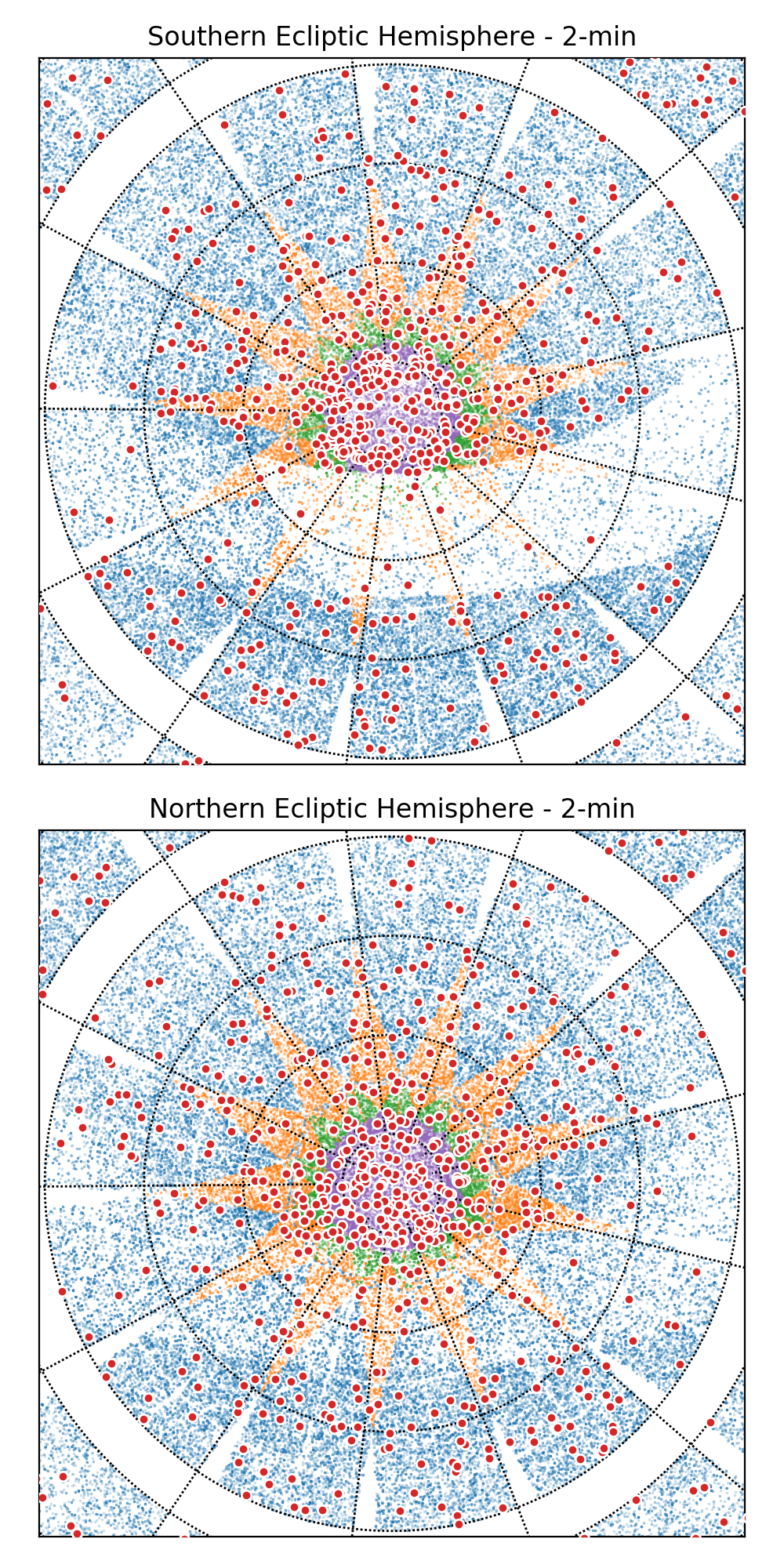}
\caption{The spatial distribution of target stars and detected planets from 2-minute cadence data. The colors of stars and planets is the same as shown in Figure~\ref{fig:skyplot-ffis}. The southern hemisphere, and to a lesser extent the northern hemisphere, has a pronounced feature of the Galactic plane running through where priorities are down-weighted because the high stellar density will dilute transit signals making them harder to detect.}
\label{fig:skyplot-2min}
\end{figure}

Our fiducial simulation has 1293 planets orbiting 2-minute cadence targets, and the 90\% confidence range of planets found in 2-minute data is 1180--1310 planets. The sky distribution is shown in Figure~\ref{fig:skyplot-2min}. There are clear differences in features between the FFI distributions and the 2-minute cadence distributions. The FFI stars are not evenly distributed, there is a lower density of stars in the southern sky. This is caused by the use of the reduced proper motion cut to identify dwarf stars, since existing proper motion catalogs are less complete below a declination of $-30^\circ$. This low density at southern latitudes is not visible in the 2-minute cadence plots because the high quality AFGK stars chosen for 2-minute cadence observations are bright enough that the proper motion catalogs are essentially complete for them. However, M dwarfs are faint enough that the proper motion catalogs are not complete for even high priority stars below a declination of $-30^\circ$, and they are undersampled among the 2-minute targets in that region.

The Galactic plane is visibly underpopulated in the 2-minute cadence data for two related reasons. Stars near the galactic plane tend to have higher flux contamination, which depressed their calculated priority. Also, photometric catalogs have a great deal of unreliability in the galactic plane in variety of ways, including proper motions, source identification, and the effects of reddening on the stellar temperatures. Therefore the priorities of all CTL stars within 15 degrees of the galactic plane were systematically down-weighted in the CTL, except for a subset of specially identified stars.

For both the 2-minute and the FFI-observed stars, we found planets more frequently closer to the ecliptic poles, where the longer observing baseline makes transit detection easier and where it is possible to find longer-period planets. 

\begin{figure}
\includegraphics[width=0.45\textwidth]{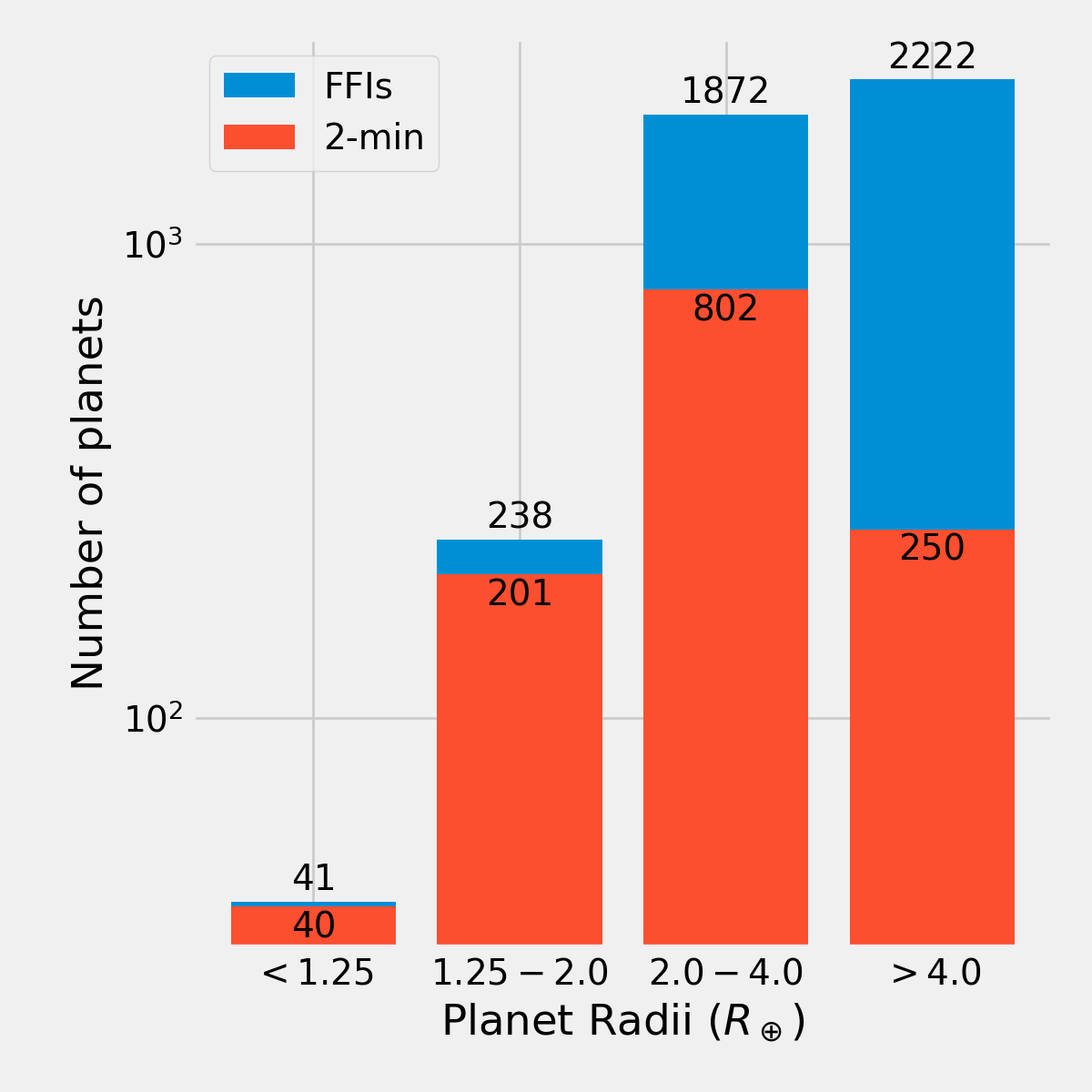}
\caption{Our simulations predict that TESS will detect a total of about 4400 planets orbiting stars on the CTL, of which 1300 will be observed at 2-minute cadence. Roughly 40 Earth-sized planets will be found, almost all of which are on the 2-minute target list. One thousand super-Earths and mini-Neptunes will also be found. Many new giant planets will be discovered, primarily through FFI data. The numbers shown above the FFI bars are total planets, and include the planets found in 2-minute cadence data.}
\label{fig:planet-radii}
\end{figure}

As shown in Figure~\ref{fig:planet-radii}, our simulation predicts that TESS will find 41 Earth-sized worlds ($<$1.25 $R_\oplus$), 238 super-Earths (1.25--2.0 $R_\oplus$), 1872 sub-Neptunes (2.0--4.0 $R_\oplus$), and 2222 giant planets ($>$4.0 $R_\oplus$) orbiting stars on the CTL. In total 279 planets smaller than 2.0 $R_\oplus$ were detected in our simulation, 90\% of which were orbiting targets observed at 2-minute cadence. The sub-Neptunes were split roughly evenly between those observed at 2-minute cadence and those found only in FFI data, but nearly 90\% of giant planets were found in the FFI data.

A summary of the properties of planets detected in FFIs and 2-minute cadence data is given in Table~\ref{tab:summary}. Full details of every planet detection in our simulation is provided in a machine readable table, with a summary shown in Table~\ref{tab:allplanets}.
%Full details of every star and planet in the simulation are provided in a machine readable table, with a sample shown in Table~\ref{tab:allplanets}.

\begin{table*}
\caption{Summary of the properties of the planets detected in our fiducial simulation. The FFI results include planets also found in the 2-minute cadence data.}
\label{tab:summary}
\begin{center}
\begin{tabular}{ l c c c| c c c} 
 \hline
  & \multicolumn{3}{c}{2-minute cadence} & \multicolumn{3}{c}{FFIs}\\
 Property& Median & 5th pctile & 95th pctile & Median & 5th pctile & 95th pctile\\
 \hline
 Host star radius ($R_\odot$)&1.02 & 0.23 & 2.44 & 1.35 & 0.32 & 3.48\\
 Host star mass ($M_\odot$)& 0.95 & 0.20 & 1.61 & 1.07 & 0.32 & 1.93\\
 Host star temperature (K)&5500 & 3200 & 7200 & 5900 & 3400 & 8000\\
 Host star brightness, $Ks$ & 9.2 & 6.7 & 11.0 & 10.0 & 7.4 & 11.5\\
 Host star brightness, TESS mag & 10.4 & 7.5 & 13.5 & 11.0 & 8.2 & 13.1\\
 Host star brightness, $V$ & 11.3 & 7.9 & 16.3 & 11.7 & 8.8 & 15.4\\
 Planet radius ($R_\oplus$)& 3.1 & 1.4 & 8.9 & 4.2 & 1.9 & 15.1\\
 Planet orbital period (days)& 8.2 & 1.7 & 34.8 & 7.0 & 1.8 & 29.0\\
 Transit duration (hours)& 3.0 & 1.0 & 8.7 & 3.9 & 1.3 & 10.4\\
 SNR & 13.6 & 7.7 & 109 & 13.3 & 7.6 & 93.7\\
 Number of transits & 7 & 2 & 65 & 6 & 2 & 51\\
 Distance (pc) & 140 & 50 & 200 & 260 & 70 & 890\\
 \hline
\end{tabular}
\end{center}
\end{table*}

\begin{table*}
\caption{Planet and host star properties for every detected planet in our simulation.}
\label{tab:allplanets}
\begin{center}
\begin{tabular}{ l l l l} 
 \hline
Num & Units & Label & Explanation\\
\hline
1 &--&TICID& TESS Input Catalog ID number of star\\
2 &deg&RAdeg& Right ascension 2000\\
3 &deg&DEdeg& Declination 2000\\
4 &deg&ELON& Ecliptic longitude\\
5 &deg&ELAT& Ecliptic latitude\\
6 &--& Priority & CTL v6.1 priority \\
7 &--&2min-target& Was this a 2-minute cadence target in our model? 1 = yes, 0 = no \\
8 &--&Camera& TESS camera number, number between 1--4\\
9 &d&Obslen& Number of days that target is observed\\
10 &--&Num-sectors& Number of sectors the target is observed for\\
11 &mag&Vmag& V-band magnitude \\
12 &mag&Kmag& Ks-band magnitude\\
13 &mag&Jmag& J-band magnitude\\
14 &mag&Tmag& TESS bandpass magnitude\\
15 &solRad&Star-radius& Stellar radius\\
16 &solMass&Star-mass& Stellar mass\\
17 &K&Star-teff& Stellar effecitve temperature\\
18 &pc&Distance& Distance of the star\\
19 &--&Subgiant& Was this star randomly selected to be a subgiant? 1 = yes, 0 = no \\
20 &--& Detected& Was this planet detected? 1 = yes, 0 = no\\
21 &--& Detected-cons& Was this planet detected using the conservative model? 1 = yes, 0 = no\\
22 &d& Planet-period& Orbital period of the planet\\
23 &Rgeo& Planet-radius& Radius of the planet\\
24 &--& Ntransits& Number of transits the planet has, 0 if planet does not transit\\
25 &--&Ars& Planet semimajor axis divided by the stellar radius\\
26 &--&Ecc& Planet orbital eccentricity\\
27 &--&Rprs& Planet radius divided by the stellar radius\\
28 &--&Impact& Planet impact parameter\\
29 &h&Duration& Planet transit duration\\
30 &ppm&Depth-obs& The observed transit depth, corrected for dilution\\
31 &--&Insol&Insolation flux the planet receives relative to that received by the Earth from the Sun.\\
32 &ppm&Noise-level& The one-hour integrated noise level of the star\\
33 &--& SNR & Combined signal-to-noise ratio of all transits, 0 if planet does not transit\\
 \hline
\end{tabular}
\tablecomments{Table 2 is published in its entirety in the machine-readable format
%available from \url{https://doi.org/10.6084/m9.figshare.6137672}
. A summary is shown here for guidance regarding its form and content.}
\end{center}
\end{table*}

About 75\% of stars were observed for a single sector. Unsurprisingly, most planets (2334, 53\%) were also only observed for a single sector and three-quarters of planets were observed for one or two sectors. Conversely, while just 2\% of CTL stars were observed for 12 or 13 sectors, 11\% of all planets detected were found around these stars. The longer observing baseline gave both higher SNR transits, and sensitivity to longer orbital period planets. The number of stars observed at 2-minute cadence for 12 or 13 sectors was fairly heavily constrained in our target selection model, therefore a relatively high fraction (60\%) of planets were found in the FFI data for the high latitude fields. Overall 70\% of planets were found only in the FFIs, but for stars that were observed between 4--11 sectors, just 40\% of planets were found only in the FFI data.

The orbital periods of our planets ranged from 0.5--99 days, which is a somewhat artificial limitation based on the occurrence rates used. The minimum orbital period of the injected transit signals was 0.5 days. While we know of several ultra-short period planets \citep[e.g.][]{Sanchis2013}, they are very rare \citep{Winn2018} and therefore will not significantly impact the planet yield. On the long period end, we simulated M-dwarf planets with periods up to 200 days, yet no planets with periods longer than 100 days were recovered, so we are confident that few long periods planets were missing here. For hotter stars, we only simulated planets with periods up to 85 days. It is likely we were missing planets orbiting stars with periods longer than 85 days. However, we only found two planets in our M-dwarf sample with periods longer than 85 days, and in the 65--85 day period range for the AFGK sample we had just 17 planets. Since the probability of a planet to transit scales inversely with orbital distance, and the number of stars with a long enough observing baseline to detect at least two transits similarly shrinks, we do not expect more than a handful of additional long period planets. We do caution that our sample should probably not be used to estimate the yield of planets showing a single transit because the 85 day limit becomes more significant. For a study of single transiting planets we point readers to \citet{Villanueva2018}.

\begin{figure}
\includegraphics[width=0.45\textwidth]{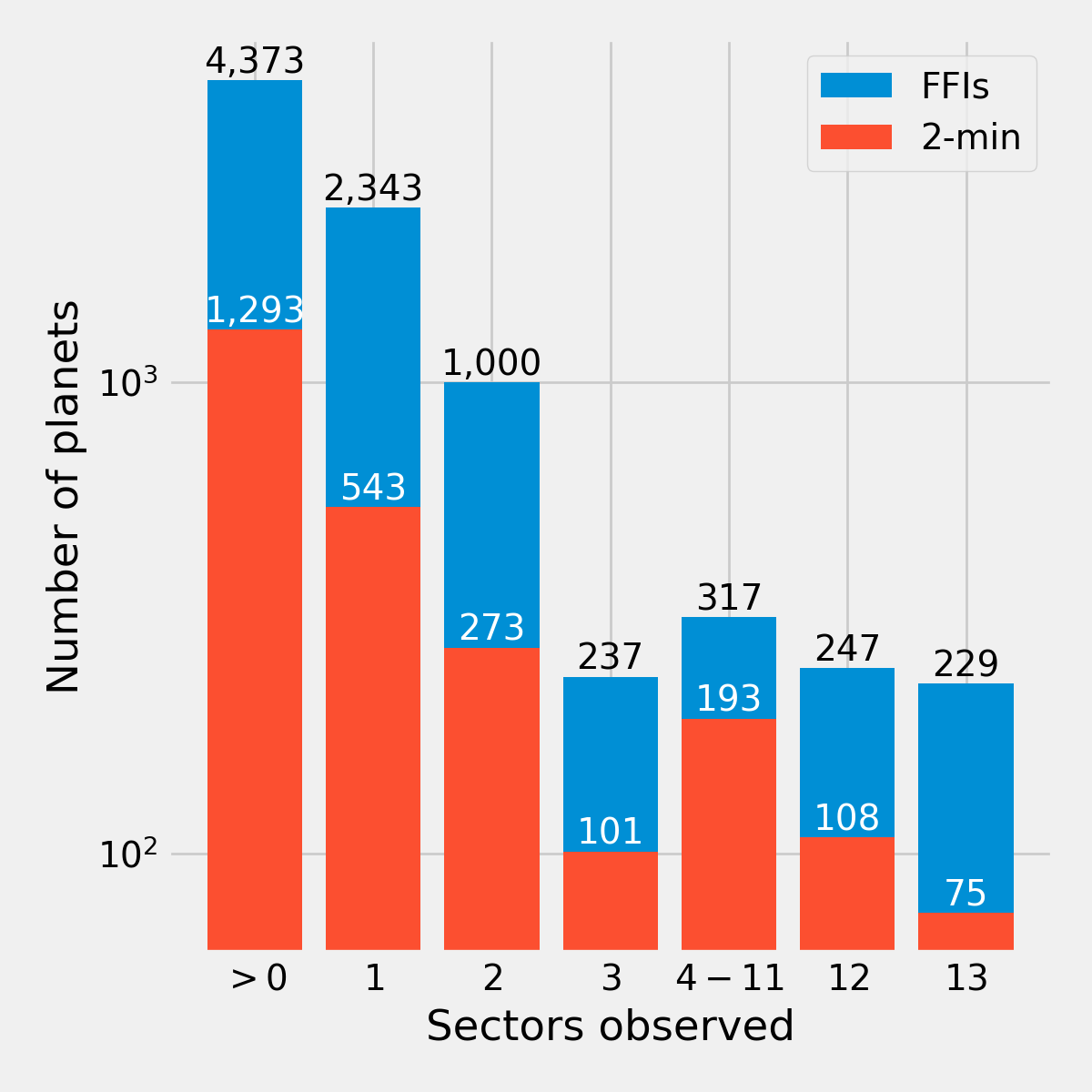}
\caption{The number of sectors that stars with detected planets were observed for, with a sector having an average observing window of 27.4 days. More than half of planets were observed for a single sector, with 10\% being observed for 12 or 13 sectors.}
\label{fig:planet-sectors}
\end{figure}

\begin{figure}
\includegraphics[width=0.45\textwidth]{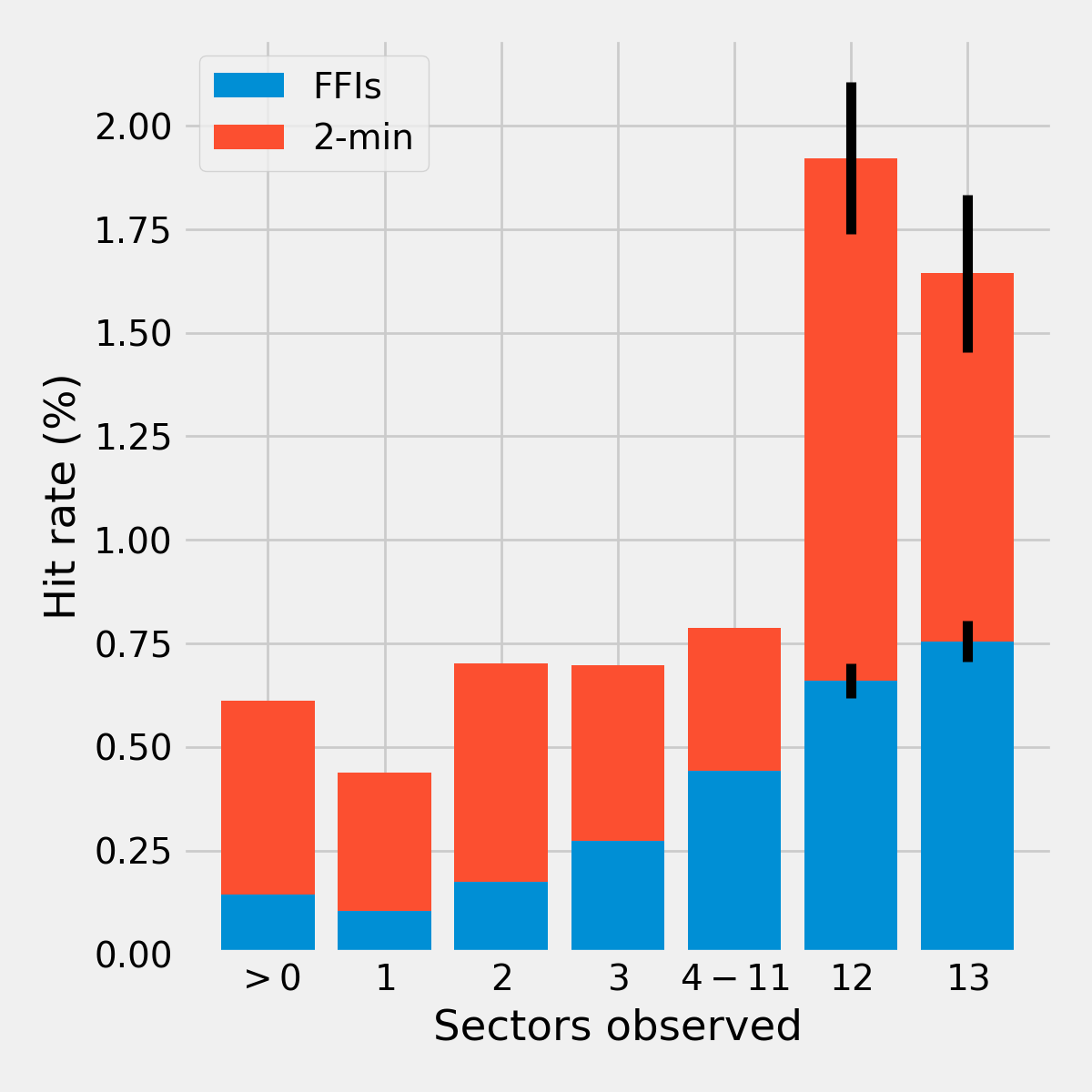}
\caption{The ratio of stars observed to planets detected as a function of the number of sectors a star is observed for. The longer a star was observed, the higher probability a planet would be detected. Targets observed at 2-minute cadence are shown in red, while blue are FFI targets. For 2-minute cadence stars the average hit-rate was 0.60\%, while including all stars on the CTL drops this to 0.14\%. While observing for a longer baseline increased the number of planets, the increase is not linear. For 2-minute cadence targets, an increase of 12x in observing baseline increased the hit-rate by a factor of just 4.4. 
There are comparatively few planets in the 12 and 13 sector bins, so we show Poisson uncertainties on these bars demonstrating that there is not a measurable difference between observing for 12 or 13 sectors. Red and blue bars are not stacked, both start at zero.}
\label{fig:hitrate-sectors}
\end{figure}

In Figure~\ref{fig:hitrate-sectors} we show the ratio of stars observed to planets detected -- which we define as the `hit rate'. Overall, the hit rate for 2-minute cadence targets was 0.60\%, while for the CTL stars not on the 2-minute cadence list the hit-rate was 0.10\%. Hit rate increases with observing duration, from 0.43\% for 2-minute cadence targets observed for 1 sector up to 1.8\% for 2-minute cadence targets with at least 12 sectors of data. %With 2-minute cadence data, we have limited target slots -- roughly 14,000 per sector. Therefore, a choice must be made at the start of each sector to observe new stars or to increase the baseline of previously observed stars. Here we show that the highest efficiency in planet detection comes from observing new stars, the hit rate divided by the number of sectors observed is highest for 1-sector at 0.44 planets-per-star-per-sector, and falls to 0.13 planets-per-star-per-sector for 13 sectors observed.

\begin{figure}
\centering
\includegraphics[width=0.45\textwidth]{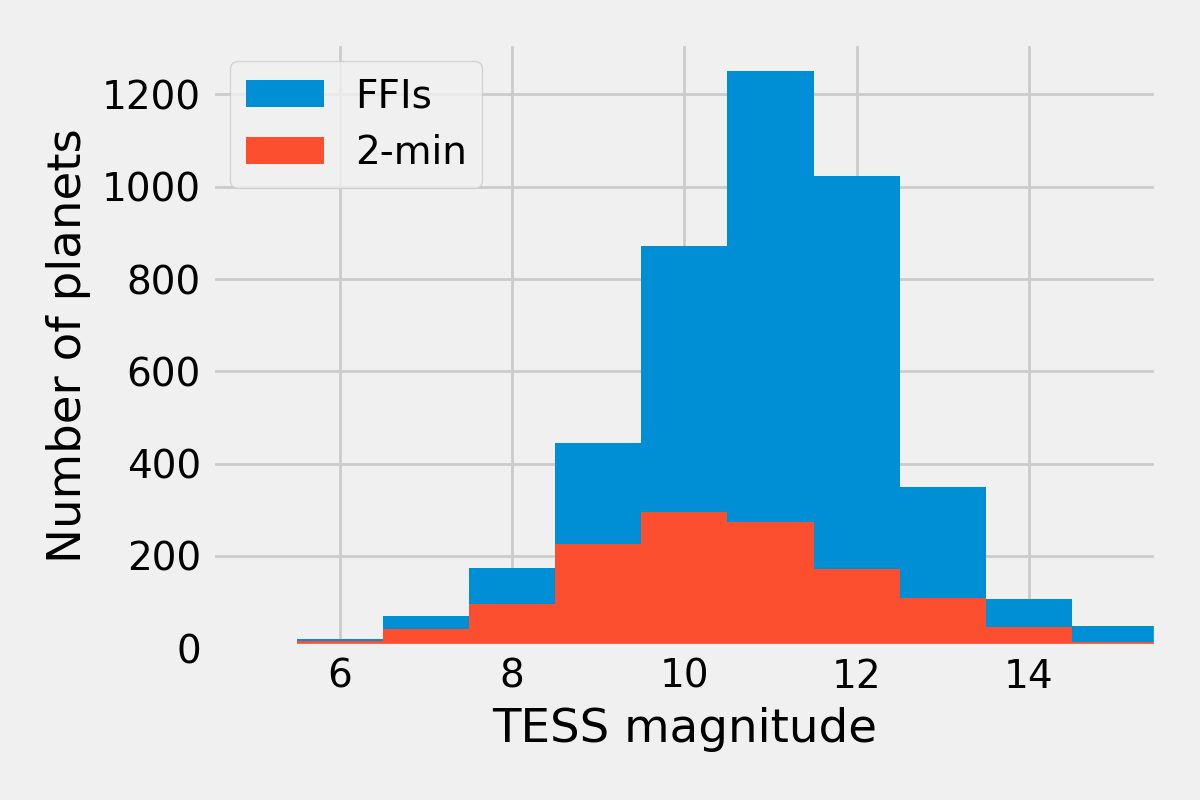}
\caption{Brightness of the planet host stars in the TESS bandpass magnitude. The median brightness of stars with planets found in 2-minute cadence data was 10.4, with a maximum range of 3.5--15.3. For planets found only in FFI data, the median brightness was 11.3, and a maximum range of 6.1--16.4. }
\label{fig:tess-mag}
\end{figure}

We found that the planet host stars range in brightness from $V$-band mag of 4.0--20, with 7 planets predicted to orbit stars brighter than 55 Cnc, currently the brightest transiting planet host \citep{Winn2011}. As shown in Figure~\ref{fig:tess-mag}, in the TESS bandpass, 90\% of planets orbited stars with magnitudes between 8.2--13.1, this compares with Kp=11.9--15.9 for Kepler planet candidates \citep{Thompson2018}. The simulated planets typically orbited stars 3 magnitudes brighter than Kepler planets. Planets around stars observed at 2-minute cadence were systematically brighter than the planets found orbiting stars observed only in FFI data, with a median TESS magnitude of 10.4 versus 11.0. 

With TESS concentrating on finding planets orbiting cool stars, it is unsurprising many planets orbited stars that were bright in the infrared. The median $Ks$-band ($\sim$2.0--2.2$\mu m$) magnitude of planets in 2-minute cadence data was 9.2 and 90\% of 2-minute cadence planets were brighter than Ks=10.7. None of the TESS 2-minute planets orbited stars fainter than the median infrared brightness of Kepler planet candidates of Ks=13.0.

\begin{figure}
\centering
\includegraphics[width=0.45\textwidth]{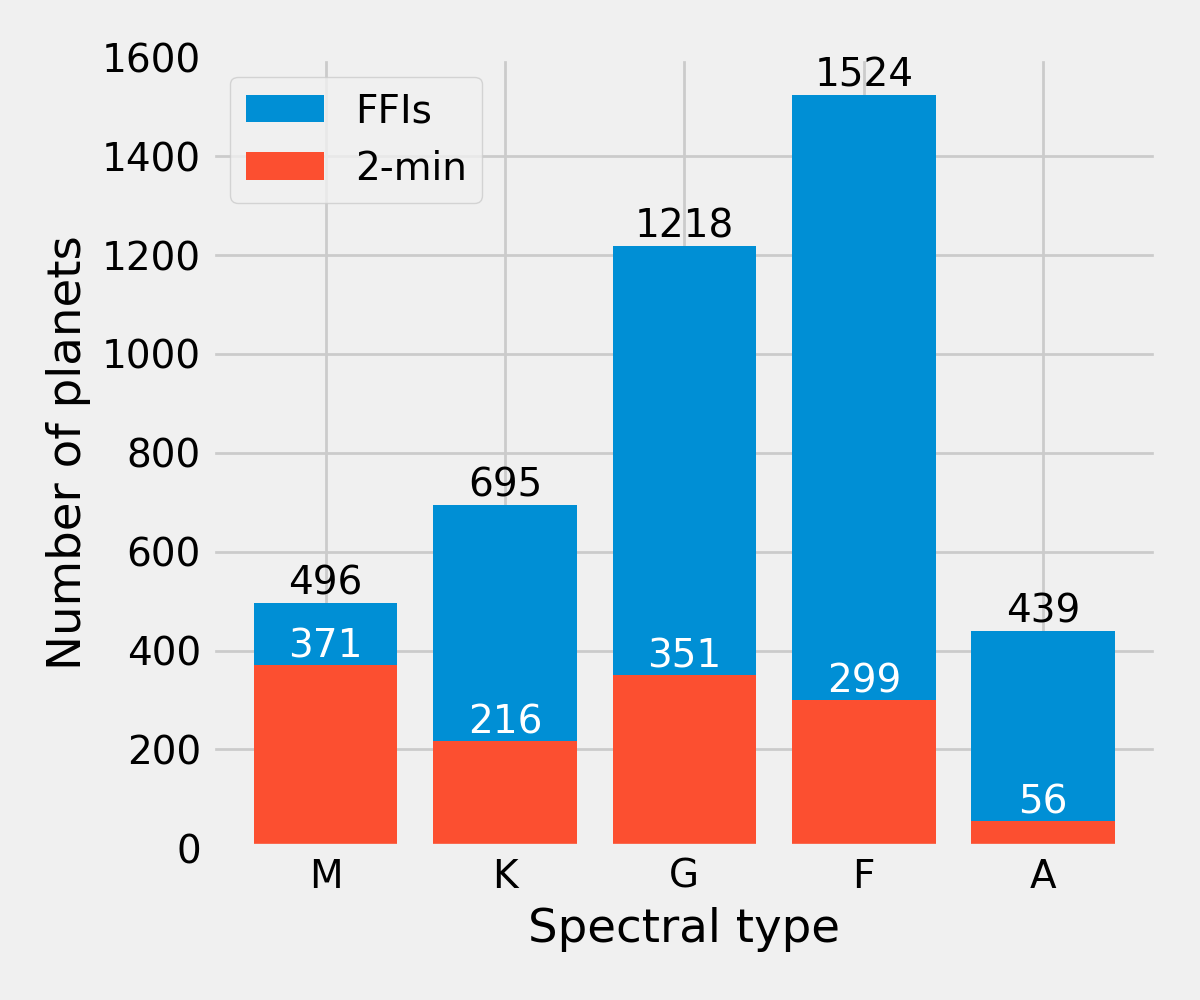}
\caption{The spectral type distribution of TESS planet-hosting stars. Our simulations predict that TESS will find 496 planets orbiting M-dwarfs, of which 371 orbit stars observed at 2-minute cadence. About half the simulated planets in 2-minute cadence data orbited stars larger than the Sun, while 80\% of planets found only in FFI data orbited stars larger than the Sun.}
\label{fig:spectral-type}
\end{figure}

The spectral type distribution of the detected planet host stars is shown in Figure~\ref{fig:spectral-type}. About a quarter of the planets found in 2-minute cadence data orbited M-dwarfs (371) with the remaining split fairly evenly between K (216), G (351), and F (299) stars. The deficit in planets orbiting K-dwarfs was caused by a deficit in K-dwarfs selected for 2-minute cadence observations. This was a result of the target prioritization strategy employed, and has been noted previously \citep{Stassun2017}. %The K-dwarf deficit (clearly visible in the lower two panels of Figure~\ref{fig:star-properties}) has been noted previously \citep{Stassun2017} and occurs because of the TESS prioritization scheme favors bright stars and small stars, with K-dwarfs not optimal in either criteria. 
A few additional planets orbiting cool stars were found in FFI data (only 125 additional M-dwarfs), but 80\% of FFI-only planets orbited stars larger than the Sun. In total about 10\% of planets in our simulated sample orbited M-dwarfs.

% \begin{figure}
% \centering
% \includegraphics[width=0.35\textwidth]{star-properties-both.png}
% \caption{Properties of the planet host stars: Ks-band ($\sim$2.0--2.2$\mu m$) (top panel), TESS bandpass magnitude (second panel), effective temperature (third panel) and stellar radius (lower panel). TESS will primarily find planets in the 2-minute cadence data around bright stars orbiting M and G dwarfs, while FFI-only planets are systematically around fainter and hotter stars, with most larger than the Sun.}
% \label{fig:star-properties}
% \end{figure}

% \begin{figure}
% \includegraphics[width=0.5\textwidth]{hit-rate-stellar-properties-both.png}
% \caption{The ratio of stars observed to planets detected as a function of stellar brightness and radius.}
% \label{fig:hit-rate-stellar-properties}
% \end{figure}

% \begin{figure}
% \includegraphics[width=0.5\textwidth]{star-properties-planet-radius-both.png}
% \caption{Ks-band brightness, host-star radius, and insolation flux, plotted against planet radius.\footnote{Most, but not all, of the very bright stars shown here (Ks$<$4) are misidentified red giant stars, which results in an incorrect star and planet radius.}}
% \label{fig:star-properties-planet-radius}
% \end{figure}

\begin{figure}
\includegraphics[width=0.45\textwidth]{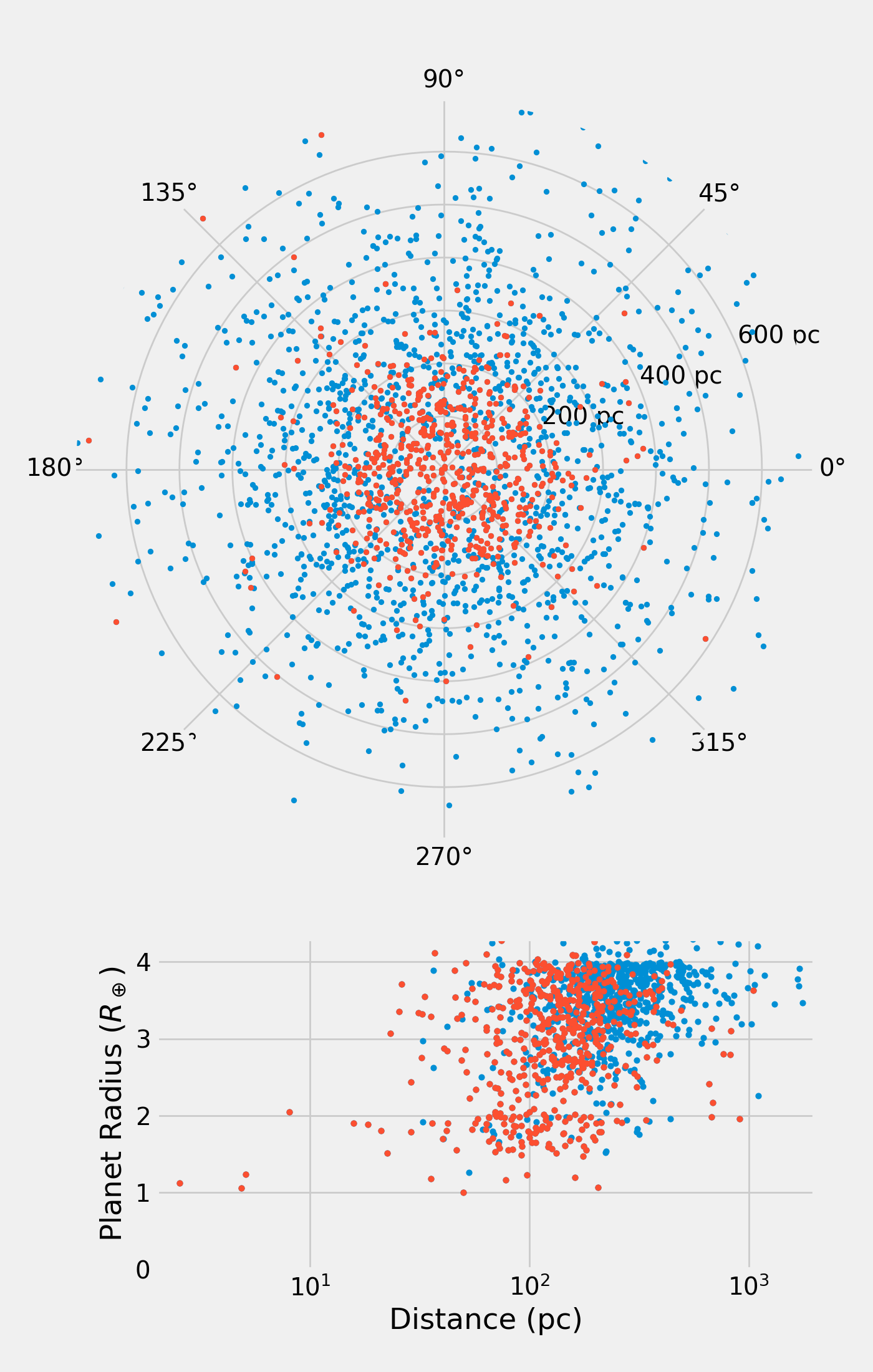}
\caption{The distances of planets found in our simulation in parsecs. The upper panel shows both distance and ecliptic latitude of the host stars, and the lower panel is distance plotted against planet radius. Almost all 2-minute cadence planets discovered by TESS will be within 300pc, with 77\% within 200pc. FFI planets were found over 1000 pc away but 90\% of planets were within 700 pc.}
\label{fig:distance}
\end{figure}

Figure~\ref{fig:distance} shows the distance to the simulated planets\footnote{Only about half of the targets in our sample had distances reported in CTL version 6.1, our statistics are based on this sample. Furthermore, a small number of the CTL reported distances were unrealistically large. These issues have been fixed in CTL v6.2.}. The closest detected planet in our simulation orbited Lalande 21185, a star 2.5 pc away. We found 46 planets within 50 pc, and 234 within 100 pc, which doubles and quadruples the number of transiting planets known within 50 and 100 pc, respectively \citep{Akeson2013}.

The circumstellar habitable zone concept has been popular since at least the 1950s \citep{Strughold1953,Shapley1953}, and refers to the spherical shell around a star where liquid water could be present on a planetary surface. \citet{Kopparapu2013} provided models for an optimistic habitable zone with boundaries of recent Venus and early Mars, which correspond to stellar fluxes of 1.78x and 0.32x the insolation Earth receives from the Sun, respectively. Our simulation contains 69 planets in the optimistic zone, of which 9 are smaller than 2 Earth-radii. All the habitable zone planets orbit M-dwarfs.

\subsection{Suitable targets for RV follow-up}
For the TESS mission to be successful, it must find planets smaller than 4 Earth-radii with a measurable radial velocity signal. We predict that TESS will find more than 2100 planets smaller than 4 Earth-radii, but many of these will orbit stars whose brightness makes follow-up challenging or impossible with current precision radial velocity facilities. While planets orbiting very faint stars have had their mass determined via radial velocity studies \citep[e.g.][]{Koppenhoefer2013}, it is typically challenging to measure masses of planets around stars fainter than V=12. We predict that TESS will find 1300 planets smaller than four Earth-radii around stars brighter than V=12. Therefore, with more than 1000 potential targets, TESS will have a plethora of targets to choose from when selecting promising RV targets. Even if just 20\% are good RV targets, this will more than triple the number of planets smaller than 4 Earth-radii with measured masses.

There are 160 planets in our sample that are smaller than 2 Earth-radii and orbit stars brighter than V=12. We currently have mass and radius constraints on fewer than 60 planets smaller than 2 Earth-radii, so TESS will potentially greatly increase this number, although the precise number will depend on whether individual stars are suitable for precise radial velocity measurements.

\subsection{Targets for atmospheric characterization}
\begin{figure}
\includegraphics[width=0.45\textwidth]{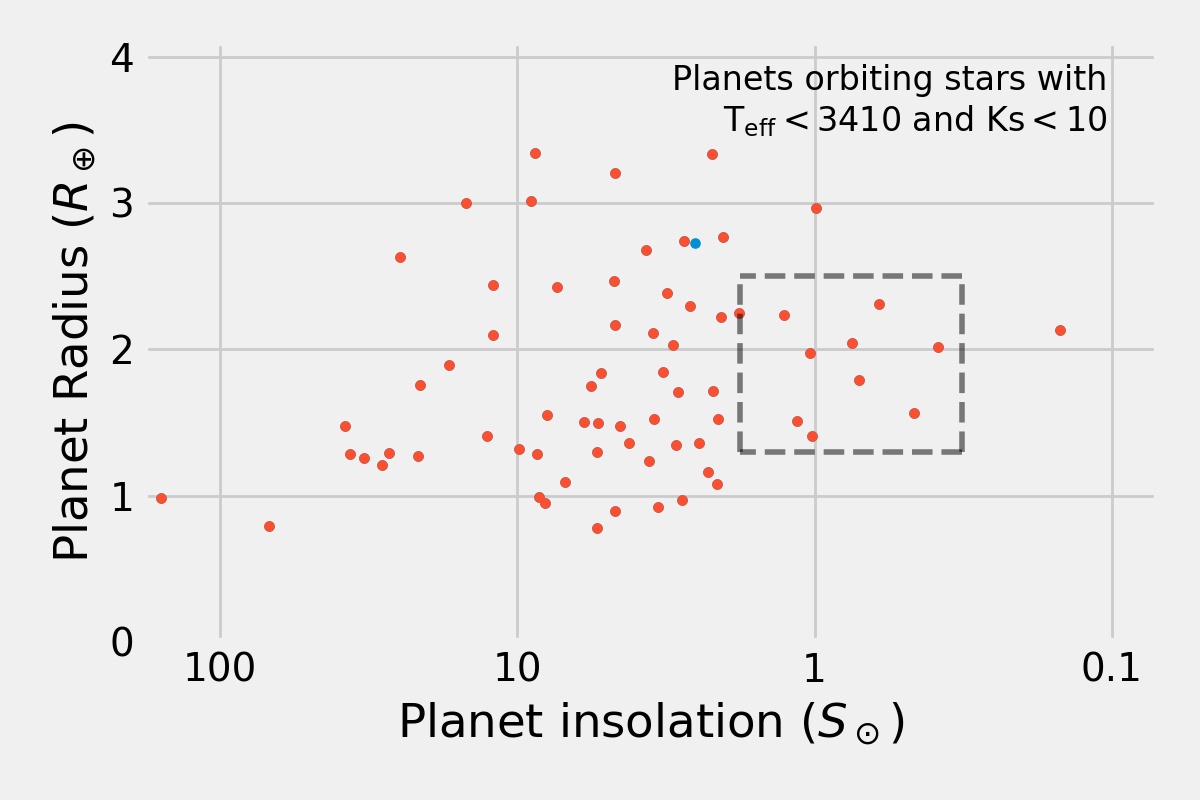}
\caption{Planets make good targets for transmission spectroscopy if they orbit bright, small stars. This plot shows planets that orbit stars with spectral type M3V or later, and that are brighter than Ks=10. The box is an approximate region showing planets that may have somewhat extended atmospheres (i.e. super-Earths) and are in the circumstellar habitable zone. There are 10 planets within this region, making up the prime JWST target sample from TESS.}
\label{fig:planet-radius-insolation-cool-stars}
\end{figure}

A second aim of the TESS mission is to find targets suitable for transmission spectroscopy using the James Webb Space Telescope (JWST). Until on-sky performance is measured, particularly the systematic noise level, there is considerable uncertainty on how JWST will perform \citep{Batalha2017}. However, we can identify the properties of planets that would make them good JWST targets using a few simple cuts. The host star should be bright in the infrared, and the star should be small. We identified simulated planets whose host stars have Ks$<$10, $T_{\rm eff}<3410 K$ which equates to M3V stars with a radius of approximately 0.37 solar-radii \citep{Pecaut2013}. In total there were 70 planets fulfilling these criteria. We show in Figure~\ref{fig:planet-radius-insolation-cool-stars} the simulated small planets we think make interesting candidate JWST targets in terms of insolation fluxes. There are ten planets in the boxed region in Figure~\ref{fig:planet-radius-insolation-cool-stars} which highlights planets that fell into the optimistic habitable zone \citep{Kopparapu2013}, and had radii between 1.25 and 2.5 Earth-radii, implying a puffed-up atmosphere \citep{Lopez2014}. These planets, along with those orbiting TRAPPIST-1 \citep{Gillon2017} and other low mass stars \citep{Greene2016,Kreidberg2016,Morley2017,Louie2018}, will form a reference sample of temperate worlds for observation by JWST.

The JWST continuous viewing zone is located within 5$^\circ$ of the ecliptic poles, and is contained within the TESS CVZ, shown in Figure~\ref{fig:sectors}. However, because of gaps between the TESS CCDs on Camera 4 (each camera is composed of a 2x2 grid of CCDs), the central 2$^\circ$ has limited coverage. In our sample we have 74 planets with ecliptic latitude $\left|b\right|>85^\circ$, of which 29 are 2-minute cadence targets and 11 are smaller than 2 Earth-radii.

% ---------------

\section{Discussion}

\subsection{Alternative selection strategies for the 2-minute cadence targets}

In addition to the nominal 2-minute cadence target selection laid out in Section~\ref{sec:starselction}, we also considered alternative strategies of selecting a higher or lower fraction of targets in the CVZ, which we call scenarios (a) and (b), respectively. There are justifications for both approaches. Placing more of the 2-minute cadence targets in the CVZ increases the overall number of 2-minute targets where TESS is sensitive to long-period planets, and potentially to smaller planets via increased SNR. On the other hand, placing more of the 2-minute targets outside the CVZ should increase the overall number of planets detected, since 13 stars can be observed in regions with single-sector coverage for each target in the CVZ.

To test these scenarios we selected targets in an identical manner to that described in Section~\ref{sec:starselction} except that in scenario (a) we included 12,000 stars in the CVZ and 2200 stars in the other cameras per sector, while in scenario (b) we select 3000 CVZ targets and 11,200 stars in the remaining cameras.

Under these two different selection strategies, we examined the number of planets found in 2-minute cadence data, compared to our nominal selection strategy. In scenario (a) we found a total of $740\pm50$ planets and in (b) we found $1380\pm60$ planets, which compares with $1250\pm70$ planets in the nominal strategy (where the reported value is the median, and uncertainties are the central 90\% of the distribution, calculated by 300 Monte Carlo simulations). These results suggest that the nominal selection strategy was reasonably successful at accomplishing the goal of maximizing the number of planets with 2-minute cadence photometry, which in turn maximizes the number of planets where we can derive precise stellar parameters through asteroseismology \citep{Campante2016}. 
Scenario (b) yielded 10\% more planets but the results were comparable within uncertainties, and the number of planets with orbital periods beyond 15 days was cut by about 10\% in scenario (b). Scenario (a) extended the tail of the orbital period distribution -- the 95th percentile shifts from 30 to 42 days -- but because of the large decrease in the total number of planets, the absolute number of long period planets was unchanged. 

In each scenario the total number of planets detected remained unchanged because almost all planets could be found equally well in 2-minute and FFI data, so the precise stellar selection had limited impact of the primary mission goals.

\subsection{A more conservative model}
\label{sec:conservative}
Our analysis so far has made two fairly optimistic assumptions, (1) that we can identify a transiting planet by observing just two transits from TESS, and (2) that we can detect all planets with a SNR$\ge$7.3. In actuality, planets with fewer than 3 observed transits are very difficult to uniquely identify using photometric survey data alone \citep[c.f][]{Thompson2018,Mullally2018}. Planets have been detected using K2 mission data \citep{Howell2014} with one \citep{Vanderburg2015} and two \citep{Crossfield2015} transits, but these cases occurred in systems where additional space-based follow-up assets were exploited or there were two other planets in the system, so the validity of the planets was less ambiguous \citep{Lissauer2012}. While with sufficient observing resources characterizing these planets is feasible to identify and confirm, they remain a challenge.
Furthermore, analyses of Kepler data have shown that using a detection threshold below 8--10$\sigma$ leads to many spurious detections \citep{Christiansen2016,Thompson2018,Mullally2018}. In K2, a threshold of SNR$>$12 was typically applied \citep{Crossfield2016} before expending follow-up resources on a candidate planet.

With these limits in mind, we took the fiducial catalog and cut planets that either had fewer than three transits, or had a combined transit SNR$<$10. This resulted in a moderate cut in the total number of planets found to 2609 total planets, of which 820 came from the 2-minute cadence data. This was a 60\% overall decrease in the total number of planets detected, but was most significant for small planets. The number of planets with radii below 2 Earth-radii decreased by a factor of two from 279 to 128 planets, with similar fractional losses in the 2--4 Earth-radii bin, but there was only a 25\% decrease in detected giant planets. 

\begin{figure}
\includegraphics[width=0.45\textwidth]{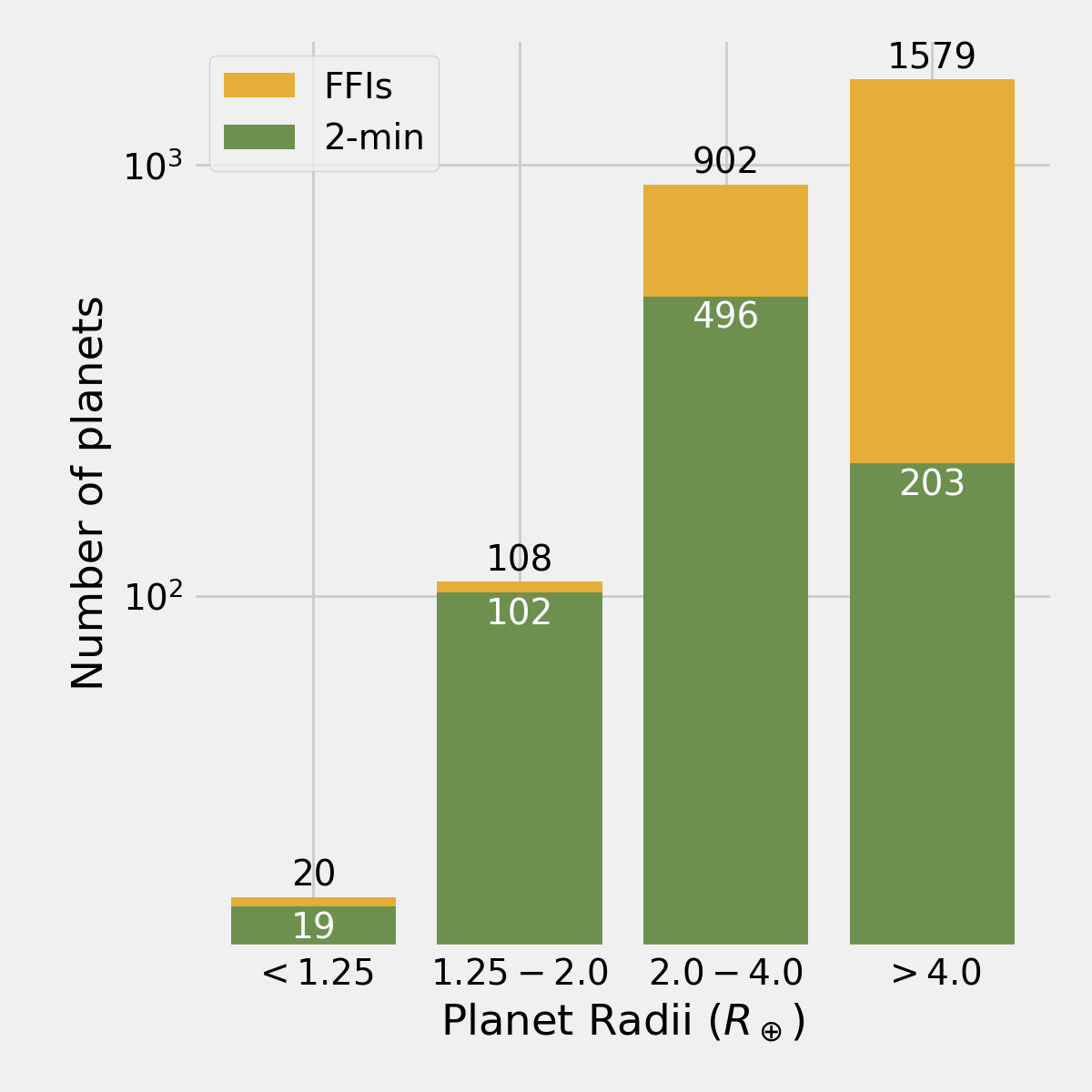}
\caption{The predicted planet radius distribution using our conservative detection model where we required at least 3 transits and a combined SNR of 10. This figure is the counterpart of Figure~\ref{fig:planet-radii}, but using our conservative detection model. The total number of planets shown is 2609, which is roughly 60\% lower than our standard detection model. This change is most signification for small planets which saw a factor of two decrease. We have intentionally changed the color scheme from previous figures to differentiate between standard and conservative models.}
\label{fig:planet-radii-conservative}
\end{figure}

The decrease in the number of planets amenable to radial velocity follow-up was roughly a factor of two, with planets smaller than 4 Earth-radii orbiting stars with V$<$12 dropping from 1312 to 616, and those smaller than 2 Earth-radii from 151 to 67. The number of habitable zone planets dropped from 69 to 28, and left just four smaller than 2 Earth-radii. The number of premium JWST targets sees a modest decrease. The number of planets orbiting stars cooler than 3410 K, with Ks$<$10 drops from 71 to 58, and the number in the dashed box in Figure~\ref{fig:planet-radius-insolation-cool-stars} dropped from 10 to 7. While these drops were significant, they are unlikely to seriously impact the primary mission goal, because there were still hundreds of small planets orbiting bright stars in the sample.

\subsection{Phantom inflated planets}
This study, and other planet yield simulations \citep[e.g.][]{Sullivan2015}, have not paid particular attention to the physical properties of giant planets, primarily because these are not a focus for the TESS mission team. Nevertheless, we are anticipating groundbreaking scientific advances in our understanding of the atmospheres of giant planets from follow-up observations of planets found by TESS -- particularly from Spitzer, HST, and JWST. As pointed out by \citet{Mayorga2018}, in the first version of this paper, there were significant numbers of giant planets that were beyond the limit of inflation for their equilibrium temperatures \citep{Thorngren2018}. The cause of this is that in the occurrence rate estimates of \cite{Fressin2013} the giant planet bin span 6--22 Earth-radii while temperate planets should rarely be larger than 12 Earth-radii. As a result of this feedback from \citet{Mayorga2018} we changed the selection function in the giant planet bins from a log-normal function to a power law. This reduced the number of phantom planets from 8\% of the total population to 1\%. We caution giant planets aficionados that there are 45 over-inflated giant planets in our simulation.

\subsection{The effects of Earth and Moon crossings}
The nature of the TESS orbit means that a subset of observations will be obscured by the Earth or Moon passing through the field of view. Cameras that receive a significant amount of scattered light from the Earth or Moon will experience larger background flux, and photometry in any camera that receives a large portion of direct light from the Earth or Moon will likely be impossible because of saturation and bleed. However, the Earth and Moon move relatively quickly through the field of view, and Earth or Moon crossings are relatively infrequent \citep{Ricker2015}. \citet{Bouma2017} estimated that the Earth and Moon will significantly affect photometric performance for 9\% of all exposures, although the lost cadences will not be evenly distributed in time or focal plane location. Camera 1, and to a less degree Camera 2, are impacted, but the effect was expected to be limited for Cameras 3 and 4. Estimating how this affects the yield is non-trivial, but we can try by using the \citeauthor{Bouma2017} estimates that 23\% of observations in Camera 1, and 12\% of observations in Camera 2, will be affected. We can then assume that the SNR of transits will scale with the square root of the number of observations, so Camera 1 targets will have 11\% lower SNR, and Camera 2 targets will have 6\% lower SNR. This causes a 13\% drop in total planets detected in our simulation, and a 9\% decrease in the number of planets orbiting 2-minute cadence targets. Early commissioning results have suggested that the effect of the Moon may be more complex than anticipated, and owing to the substantial uncertainty in the impact of Earth and Moon crossings, we have not included Earth and Moon crossings in our yield statistics.

% \subsection{Alternative prioritization schemes}
\subsection{Astrophysical false positives}
\label{sec:falsepositives}
\citet{Sullivan2015} performed a careful analysis of the sources and rates of false positives expected in the TESS 2-minute cadence data, and we have not reproduced that work here. They estimated that TESS will find over 1000 astrophysical false positives in 2-minute cadence data, but described promising mitigation strategies that utilize follow-up observations and statistical methods to reduce this by a factor of 4 or more.

The ratio of false positives to detected planets will not be uniform over all stars observed by TESS, but will vary as a function of hit-rate. In Section~\ref{sec:results} we showed that the hit-rate for 2-minute cadence targets is %a factor 4 higher than for all stars, and 
a factor of 5.5 higher than FFI-only stars. Assuming each star has the same chance of yielding a detection of an astrophysical false positive, the fraction of true planets found to false positives will be lower for the FFI-only detections than for 2-minute cadence targets. The reason is that fewer planets are found per stars observed but the same number of false positives are detected.
Using the false positive rate from \citet{Sullivan2015} of 1 false positive per 180 stars observed yields one astrophysical false positive per planet detection. However, for the FFI-only targets the ratio of false positives to planets detected increases to more than five per true planet discovered.%{\bf[I'm not sure I follow this paragraph]} I've tried to make it clearer, let me know if it's still confusing - Tom.

Furthermore, stars on the CTL that are not included in our 2-minute cadence sample are, on average, 2 magnitudes fainter than the 200,000 stars observed at 2-minute cadence. This means that mitigation strategies that rely on follow-up observations will be significantly more challenging. Given essentially all small planets will be found in the 2-minute cadence data, only the most intrepid of exoplaneteers will want to commit significant resources to discovering and following-up planets in FFI data.

\subsection{Planets detected around stars not in the CTL}
\label{sec:excluded-stars}
In Section~\ref{sec:starselction} we simulated planets orbiting stars that are in CTL version 6.1. This totals roughly 3.2 million stars, but includes only those stars that the TESS Target Selection Working Group considered as potential 2-minute cadence targets. The limited number of slots available for 2-minute cadence requires a careful consideration not just of the overall potential for planet detections around a given star, but also comparison of the relative planet detection potential between stars, along with the scientific value of the resulting planets. The CTL was constructed to permit a quantitative relative ranking of the best stars to select for the 2-minute cadence slots, not to identify all stars with detectable planets. While in this work we have adopted the set of several million stars in the CTL as the primary sample to investigate, stars not in the CTL might also yield some planet detections in the FFI data. The reason we adopted this approach is the same reason for the construction of the CTL in the first place -- our analysis of planet yield among a population of several million stars is much more tractable than conducting the analysis for all 470 million stars in TIC-6.

Explicitly removed from the CTL are stars with a reduced proper motion that flags them as giants, stars with parallax or other information that flags them as giants or subgiants, dwarf stars that are somewhat hot and relatively faint but not as faint as some dwarf stars that are included, and faint dwarf stars. The magnitude cut used in the CTL is TESS magnitude of 12 for stars hotter than 5500 K, and TESS magnitude 13 for cooler stars, although faint cool dwarfs are explicitly included via a specially curated target list \citep{Muirhead2018}. The CTL therefore generally excludes hot stars, faint stars, and evolved stars, in favor of bright, cool dwarfs.

Only a handful of transiting planets have been detected around red giants \citep[e.g.][]{Burrows2000,Huber2013,Barclay2015,Vaneylen2016,Grunblatt2016,Grunblatt2017} because finding these planets is extremely challenging. Transit depth scales with the square of the stellar radius, so planets orbiting large stars are hard to find. Therefore, the frequency of planets orbiting giant stars is relatively poorly constrained. However, TESS will observe hundreds of thousands of red giants brighter than 11th magnitude in the TESS bandpass \citep{Huber2017a} and will certainly detect planets orbiting these stars. However, Kepler observed roughly 16,000 red giants \citep{Yu2018} and found only a handful of planets. With a factor 20 or so increase in the number of red giants from TESS, we might expect of order 100 new planets. This estimate is comparable to that of \citet{Campante2016}, who perform a much more careful analysis and predicted that TESS will find roughly 50 planets orbiting red giants.

The brightness cuts applied to the TIC in creating the CTL have a larger impact on our yield estimates. At 12th magnitude the TESS 1-hour integrated noise level is 600 ppm. This equates to detecting a Neptune-size planet with three transits around a solar radius star, while at 13th mag the noise is 1200 ppm which is equivalent to a 6 Earth-radii planet. So it is certainly the case that many stars not included in the CTL may have planets detectable with TESS. To detect a Jupiter with three transits around a Sun-like star would require a maximum 1-hour integrated noise of approximately 4000 ppm which corresponds to a TESS magnitude of 14.7. The TIC lists 16.0M stars with temperatures above 5500 K, log$g$ above 3.9, and TESS magnitude of 12--14.7, and 4.2M with temperature between 4000--5500 K, $\log{g}$ above 4.2, and brightness between 13--14.7 (where we cut at 4000 K because the cooler stars are included via the cool star curated list). In our fiducial sample, the frequency of detected planets larger than 4 Earth-radii was 0.069\%. Assuming an equal detection rate for fainter stars in the 4+ Earth-radii bin as for brighter stars we would expect to find 14,000 additional giant planets. Even under our conservative model, the rate is 0.050\%, or 10,000 additional planets.

While these planets will appear in the FFI data, they are not prime targets, hence their exclusion from the CTL, because the planets will be hard to detect and harder to follow up and confirm owing to their faintness and higher crowding. Using the logic described in Section~\ref{sec:falsepositives}, the astrophysical false positive rate in this part of the parameter space is also very high. With a hit-rate around 0.05\% and a false positive rate likely to be comparable to that found by \citet{Sullivan2015} of 1 per 180 stars observed, we expect a factor of more than 11-to-1 false positive to true planets detected. Thus we caution that searching for planets in this regime is fraught with challenges.

\begin{figure}
\includegraphics[width=0.45\textwidth]{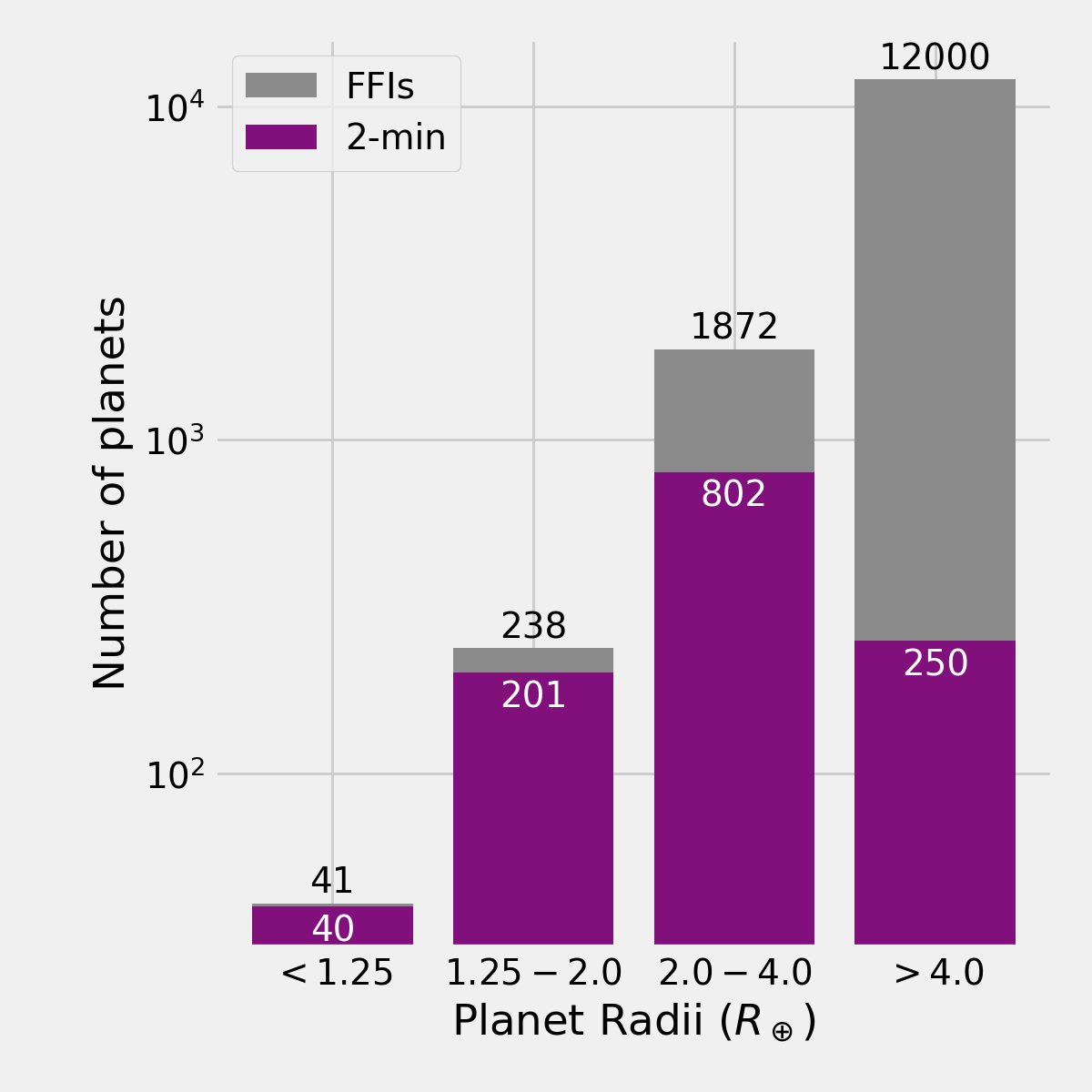}
\caption{
The predicted planet radius distribution including large planets orbiting faint stars outside of the CTL. The total number of planets that we predict TESS could find is up to 14,000. This figure is the same as Figure~\ref{fig:planet-radii} but includes the additional large planets orbiting faint stars. We have intentionally changed the color scheme from previous figures to differentiate from our simulated yield.
}
\label{fig:planet-radii-giant-planets}
\end{figure}

The omission of these potential host stars from our analysis leads to a large underestimate in the overall planet yield of the mission, although that is almost entirely in the giant planet regime. In Figure~\ref{fig:planet-radii-giant-planets} we show our final distribution of planet radii and include the sample of giant planets orbiting faint stars, using the conservative yield estimate. This results in a total planet yield of 14,000 transiting planets. However, as discussed, these planets will be resource intensive both to confirm and to meaningfully analyze.

One further source of additional planets is from M-dwarfs in the Southern Hemisphere. As mentioned in Section~\ref{sec:starselction}, there is a deficit of cool stars below $-30^\circ$ declination, caused primarily by the lower completion of proper motion catalogs where northern hemisphere telescopes are unable to observe. This manifests in fewer planets detected around cool stars in the south. In the 2-minute cadence data, there are 2.6x as many planets orbiting stars cooler than 3900 K north of declination $30^\circ$ than south of declination $-30^\circ$. Including the FFI planets, this increases to 3x as many northern as southern planets (233 versus 74 planets). With GAIA data release 2 now available, it is probable that new M-dwarfs in the south will be identified. This will help to recover additional planets orbiting cool stars not identified as dwarfs in the CTL. Given that this could potentially yield new candidate planets for JWST there is a pressing need for this work.

\subsection{Comparisons with earlier estimates}
\citet{Sullivan2015}, \citet{Bouma2017}, and \citet{Ballard2018} have previously estimated the planet yield from TESS. These previous studies selected stars from a simulated Galactic model rather than real stars, and therefore we expect there are moderate differences between our predicted yields and previous studies. Additionally, we used different selection strategies for both 2-minute cadence targets and for FFI stars. We built a realistic 2-minute cadence star selection model that limits the stars observed at the pole cameras to just 6,000 stars per hemisphere, whereas the previous works assumed TESS can observe many more stars in the CVZ than is possible with the flight hardware configuration used. We also use a different prioritization metric than previous work, which is based on the metric used by the TESS Target Selection Working Group. For the FFI targets we primarily consider those within the CTL, whereas different cuts on brightness are made in earlier works. Therefore, we expect to see significant differences in the planet yield for giant planets.

\citeauthor{Sullivan2015} %is the mission team's paper and 
predicted 1700 planets in 2-minute cadence data, of which 560 are smaller than 2 Earth-radii. \citeauthor{Bouma2017} used the same methodology and software as \citeauthor{Sullivan2015}, but fixed a number of software bugs and modified a number of parameters. They also predicted 1700 planets from 2-minute cadence data, of which 430 were smaller than 2 Earth-radii. The total 2-minute cadence planet yield in both these studies was about 30\% larger than we have predicted, but the number of planets smaller than 2 Earth-radii in our study is lower by a factor of 1.7 and 2.3 than \citeauthor{Bouma2017} and \citeauthor{Sullivan2015}, respectively. However, given the different selection strategies, it may be more reasonable to compare the combined 2-minute cadence and FFI yields. Where \citet{Bouma2017} and \citet{Sullivan2015} differ is in their star selection for FFI targets. \citeauthor{Bouma2017} limit their selection to the top ranked 3.8M stars using a similar priority metric to the one applied in CTL 6.1. This enables easy comparison with our 3.2M star sample. On the other hand, \citet{Sullivan2015} consider all stars brighter than $Ks=15$ totalling 150M stars, which we can compare with our analysis in Section~\ref{sec:excluded-stars}.

Our total simulated yield is remarkably similar to \citeauthor{Bouma2017}, with 41 versus 49 Earth-sized planets, 238 versus 390 super-Earths, 1900 versus 2000 mini-Neptunes, and 2200 versus 2500 giant planets, for this work and \citeauthor{Bouma2017} respectively. The only area where we see a significant deviation is for super-Earths, which we attribute to differences between the Galactic model and real stars.

Compared to \citet{Sullivan2015,Sullivan2017}, we predict lower totals in all bins. However, as mentioned by \citeauthor{Bouma2017}, the number of Earths and super-Earths is overestimated by around 30\% owing to a bug in their calculation of the dilution from background stars. Taking this into account, our number of Earths matches both \citeauthor{Bouma2017} and \citeauthor{Sullivan2015}, while the super-Earths are comparable. Our rate of giant planets predicted in Section~\ref{sec:excluded-stars} is consistent with \citeauthor{Sullivan2015} %The while the deviation between this work and \citeauthor{Bouma2017}, and \citeauthor{Sullivan2015} can be attributed to 3--4 Earth-radius planets orbiting faint K-dwarfs.

\citeauthor{Ballard2018} used the framework and detection rates of \citet{Sullivan2015}, but focus entirely on M1--M4 dwarfs, and made significant changes to the occurrence rates to account for covariances between planets in the same systems. In comparison, our analysis of the M-dwarf population is simplistic. \citeauthor{Ballard2018} predicted a 50\% increase in the rate of planets orbiting these cool stars compared to the occurrence rates used by \citeauthor{Sullivan2015} (and this work). They predicted $990\pm350$ planets around M1--M4 stars, while we predicted 410 planets orbiting stars with temperatures of 3100--3800 K. If the \citeauthor{Ballard2018} occurrence rate has a similar impact to our yields as it had on \citeauthor{Sullivan2015}, and given comparable yields between our studies, we would expect an additional 50\% planets in this parameter space, which is 200 more planets orbiting cool stars. Assuming the increase is uniform in planet size, we might expect an increased yield that includes 14 additional Earths, 42 additional super-Earths, and 142 additional mini-Neptunes. The yield could be even higher if we are able to identify additional M-dwarfs in the southern sky, as discussed in Section~\ref{sec:excluded-stars}.
%we think this is probable, 

% \subsection{Assumptions and caveats}
% every star has the same PSF

\section{Conclusions}
The TESS mission will find a large number of transiting planets. However, up until recently the number and physical properties of the planets that will be discovered has been estimated using simulations performed before the TESS observing strategy, 2-minute target list, and flight hardware had been finalized. Here we simulated TESS detections of transiting planets using the CTL for our star selection. We have estimated that TESS will find more than 14,000 exoplanets, of which $4400\pm110$ orbit stars in the CTL and $1250\pm70$ will be observed at 2-minute cadence. TESS will find over 2100 planets smaller than 4 Earth-radii, of which 280 will be smaller than 2 Earth-radii.

The key design feature that distinguishes TESS from Kepler is that it will observe brighter stars, emphasizing finding planets that can be followed up more readily from the ground. TESS planets range in V-band brightness from 4--20, with 80\% of predicted planets orbiting stars brighter than V=13.0. Assuming V=12 as the limit for recovery of a mass via precision radial velocity observations, we predict that TESS will have a sample of 2500 planets for radial velocity observations, of which 1300 will be smaller than 4 Earth-radii, and 150 smaller than 2 Earth-radii. This will provide a plethora of planets to characterize; the TESS follow-up observers should have little problem meeting mission requirements of measuring the masses of 50 planets smaller than 4 Earth-radii.
We predict that TESS will find 7 planets orbiting stars brighter than 55 Cnc, the brightest transiting planet host.

There is significant interest in finding habitable zone planets from TESS. We predict around 70 habitable zone planets will be detected and all will orbit M-dwarfs, with 9 habitable zone planets in our simulations with radii smaller than twice that of Earth's.
Our simulations predict that TESS will find 70 planets orbiting bright mid-M-dwarfs (Ks$<$10, M3V or later), 10 of which fall into the optimistic habitable zone, making them prime JWST targets. 

We have shown that nearly all planets valuable for contributing to mission goals related to radial velocity and JWST targets will be found in 2-minute cadence data. This is to the great credit of the teams that worked to create the CTL. The availability of 2-minute cadence data will permit more accurate measurements of the radii and orbital configurations of the detected planets. We explored how target selection choices affect the target yield and find that the distribution of targets between the CVZ and shorter observing baseline is well balanced between collecting 2-minute cadence data for the maximum number of planets, and finding long period planets.

There are a large number of stars that are not in the CTL that might host a detectable planet. These stars were intentionally not included in the CTL, and for good reason. They are unlikely to host detectable small planets, and any planets found will be hard to follow up. While there may be as many as 10,000 additional giant planets around the faint stars in the TESS data, we have shown that the astrophysical false positive rate might be as high as 11 false positives per true planet, and there may be as few as one planet detected per 2000 stars searched. While less severe, we anticipate a high astrophysical false positive rate for stars on the CTL but not included in the 2-minute cadence sample because the ratio of detected planets to stars observed is five times lower than for stars observed at 2-minute cadence. %Lastly, if the FFI light curves are of comparable photometric quality to the 2-minute cadence light curves, planets orbiting larger stars will have less need of high-cadence light curves, since the longer transit durations will generally allow adequate sampling of the transits.

The mission's target of finding planets with SNR$\ge$7.3 and only two transits may be overly aggressive, based upon experience with Kepler and K2 data. We explored an alternative model that applied more conservative detection thresholds of SNR$\ge$10, and requiring three transits. This results in a decrease in the yield estimate of approximately 50\% for planets smaller than 4 Earth-radii, and occurs across all parameter spaces considered. However, even if this conservative model is realized, more than enough planets will be found to ensure mission success.

This work builds upon studies by \cite{Sullivan2015} and \cite{Bouma2017}, and would not be possible without their efforts. We do see a moderate decrease from previous yields estimates, although our numbers are remarkably similar to those \citeauthor{Bouma2017} presented, considering the different stellar selection strategies.

It will not be long before TESS planets are discovered. The real excitement will come from learning about these new worlds using data from ground and space-based facilities. The legacy of TESS will be a catalog of the planets that will be the touchstone planets for years to come. TESS will discover which of our nearest stellar neighbors have transiting planets. The brightest host star in our simulation is 70 Oph A, where we recovered a simulated Earth-sized planet. Were this simulation real, on a clear night from a dark site we could point to this star and tell our friends, ``that star there has a planet.''

\acknowledgments{
This work has been made possible though the valiant efforts of the TESS Target Selection Working Group. Without their dedicated effort to create such a high quality catalog this work would not be possible.
We thank Luke Bouma, Chelsea Huang, Joshua Schlieder, Daniel Huber, Scott Gaudi, Diana Dragomir, Steven Villanueva, Laura Mayorga, Careloine Morley, Laura Kreidberg, and Dana Louie for insightful discussions that greatly improved the manuscript. We also want to recognize the TESS team at MIT for their tireless work in making the mission happen.
This research has made use of the NASA Exoplanet Archive, which is operated by the California Institute of Technology, under contract with the National Aeronautics and Space Administration under the Exoplanet Exploration Program.
This work has made use of the CTL, through the TESS Science Office’s target selection working group (architects K. Stassun, J. Pepper, N. De Lee, M. Paegert, R. Oelkers). The Filtergraph data portal system is trademarked by Vanderbilt University.
}

\software{Matplotlib \citep{matplotlib}, 
SciPy \citep{scipy},
NumPy \citep{numpy},
IPython \citep{ipython},
Jupyter \citep{jupyer},
Pandas \citep{pandas},
Astropy \citep{astropy},
Astroquery \citep{astroquery},
tvguide \citep{Mukai2017},
coco \citep{coco}
          }

\appendix
\section{Planet radius as a function of distance}
\begin{figure}
\includegraphics[width=\textwidth]{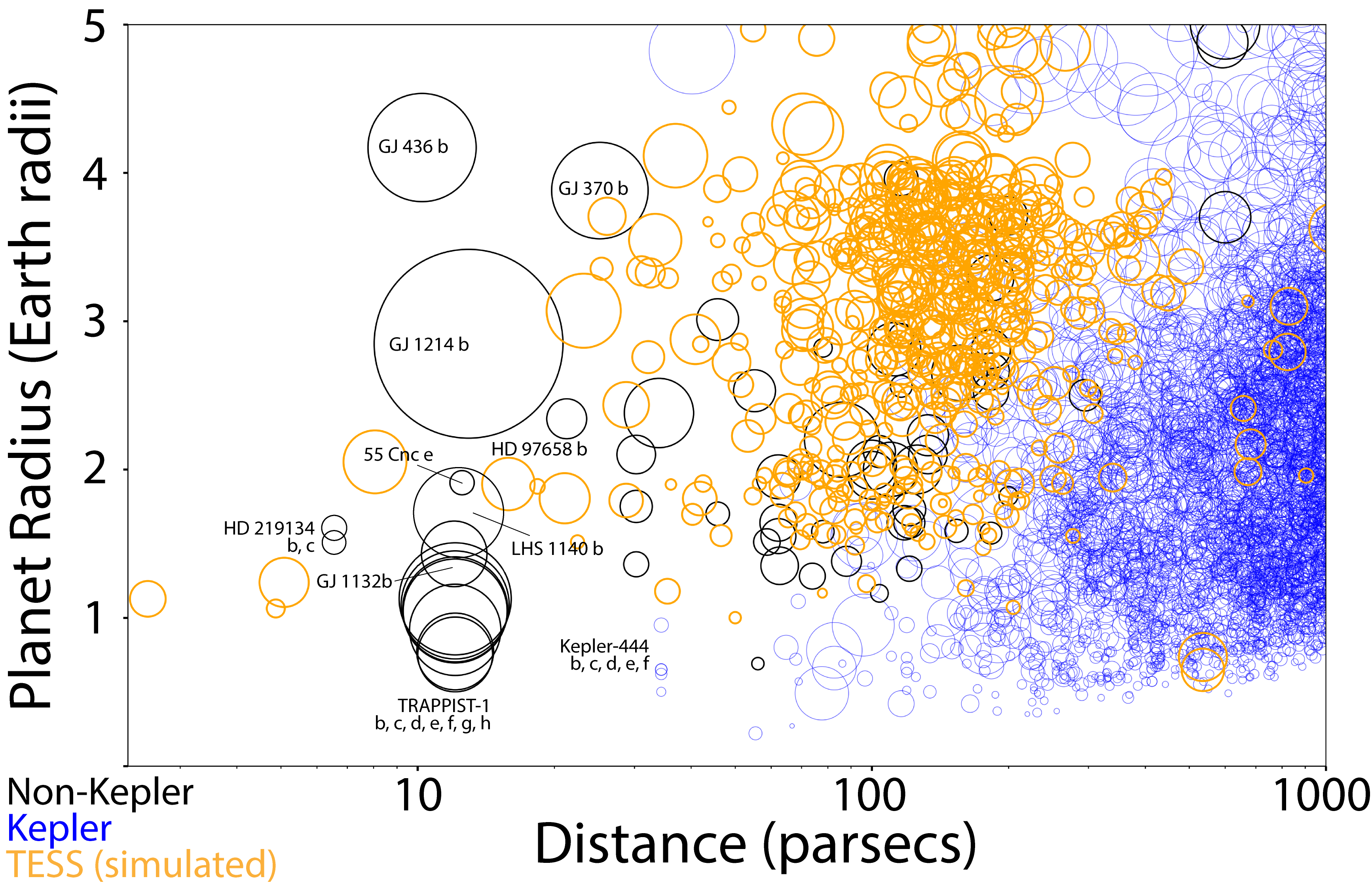}
\caption{Orbital distance versus planet radii. This plot updates a widely shared figure created by Z. Berta-Thompson, to now include our new simulation results. Kepler planet candidates from \citet{Thompson2018} are shown in blue, our simulated 2-minute cadence detections in orange, and planets detected using other telescopes in black. The size of the circle is proportional to the transit depth. A subset of nearby planets are marked. Data was extracted from the Exoplanet Archive \citep{Akeson2013}. Three planets in our simulation orbit stars closer than the nearest known transiting planet system HD 219134.
}
\label{fig:zachplot}
\end{figure}

Zach Berta-Thompson created a figure using data from \citet{Sullivan2015} that has been widely shared because it is both informative of TESS' capabilities and aesthetically pleasing. We have reproduced Berta-Thompson's plot in Figure~\ref{fig:zachplot}, with our revised TESS yield estimates.

\end{document}